# The observation of bulk superconductivity in Rhombohedral $ReO_3$ under pressure

S. Huyan [1,2*†], R. F. S. Penacchio[1,3*], L.- L. Wang[1,2*], J. Schmidt[1,2‡], D. Zhang[4], B. Lavina[4,5], Z. Li[1,2], R. A. Ribeiro[1,2], T. J. Slade[1,2], J. Zhao[5], S. L. Morelhão[3], P. C. Canfield[1,2], S. L. Bud'ko[1,2†]

[1] *Ames National Laboratory, US DOE, Iowa State University, Ames, Iowa 50011, USA*
[2] *Department of Physics and Astronomy, Iowa State University, Ames, Iowa 50011, USA*
[3] *Institute of Physics, University of São Paulo, São Paulo, SP, Brazil*
[4] *Center for Advanced Radiation Sources, The University of Chicago, Chicago, Illinois 60637, USA*
[5]*Advanced Photon Source, Argonne National Laboratory, Argonne, IL 60439, USA*

[*] These authors contributed equally.

[†] shuyan@iastate.edu, budko@ameslab.gov.

Understanding how lattice geometry enables superconductivity in oxides remains a central challenge. Here, we report a systematic study of $ReO_3$ up to 80 GPa. Synchrotron x-ray diffraction, Raman spectroscopy, electrical transport, dc magnetic susceptibility, and first-principles calculations establish a sequence of pressure-induced structural transitions, from cubic $Pm\bar{3}m$ to $Im\bar{3}$ followed by the emergence of a rhombohedral $R\bar{3}c$ phase accompanied by bulk superconductivity with a maximum $T_{c,\ onset}$ ~17.5 K. DC magnetic susceptibility and trapped-flux magnetization measurements demonstrate that bulk superconductivity is confined to the pressure range where $R\bar{3}c$ phase is dominant. Density functional theory calculations show strong electron-phonon coupling in the $hR24$-$R\bar{3}c$ structure, with substantial contributions from both low-frequency Re vibrations and high-frequency oxygen-related phonon modes, yielding a calculated $T_c$ comparable with the experiment. Upon further compression above ~35-40 GPa, powder X-ray diffraction results indicate a symmetry-lowering structural transition. Whereas the experimental diffraction patterns can be best described by a rhombohedral-derived $R32$-like average distortion with effective enlargement of the crystallographic unit cell, enthalpy calculations identify a lower-symmetry $mP16$-$P2/c$ structure driven by phonon instability of the $R\bar{3}c$ phase. This reconstructed higher-coordination phase has a reduced density of states at the Fermi level, weaker electron-phonon coupling, and a much lower calculated $T_c$, providing a microscopic explanation for the loss of bulk superconductivity in the higher-pressure phase. These results show that bulk superconductivity is stabilized within the rhombohedral structure, where pressure-induced lattice reconstruction supports enhanced electron-phonon coupling through cooperative Re-O lattice dynamics.

[‡] Currently at Instituto de Física de Buenos Aires, CONICET-Universidad de Buenos Aires, Pabellón 1, Ciudad Universitaria, CABA, 1428, Argentina.

## Introduction

Rhenium trioxide ($ReO_3$) is one of the deceptively simple $ABX_3$ perovskites with unoccupied A-cation site, where strong Re-5$d$–O-2$p$ hybridization leads to textbook lattice dynamics and unusual transport properties [1]. Corner-sharing $ReO_6$ octahedra support rigid-unit tilts and a soft $M_3$ phonon mode that underpins its negative thermal expansion (NTE) and tilt-driven pressure-induced structural transitions [2,3]. At ambient pressure, $ReO_3$ is in the cubic $Pm\bar{3}m$ structure. Recent pressure dependent neutron/X-ray/Raman work shows that above ~ 0.7 GPa a cubic $Im\bar{3}$ phase is stable to at least ~15 GPa and that several earlier noncubic assignments arose from high-flux X-ray-induced artifacts and/or non-hydrostaticity [4] - a methodological point that is crucial for interpreting emergent states in compressed $ReO_3$.

At ambient pressure, in the $Pm\bar{3}m$ structure, $ReO_3$ remains non-superconducting down to 20 mK and has a relatively weak EPC constant of $\lambda \approx 0.3$ [4]. This behavior contrasts with the pressure-stabilized rhombohedral phase, in which structural reconstruction substantially modifies both the electronic structure and phonon-mediated coupling. In addition, $ReO_3$ displays extremely large magnetoresistance (XMR) governed by perfect carrier compensation for $H \parallel c$ and by open-orbit trajectories near $\theta \approx 15°$ [6], underscoring the primacy of Fermi-surface geometry to magnetotransport properties, a transport regime that favors high mobility but not SC, such as was found recently in other binary compounds in $WTe_2$ [7], LaBi [8], $LaSb_2$ [9], and $Re_3Ge_7$ [10]. Based on these observations, we suggest that the compression could rearrange the oxygen lattice of $ReO_3$ and thus boosting EPC and enabling the emergence of a SC dome in the high-pressure rhombohedral phase (while the XMR fingerprints weaken).

Beyond these observations, Shan et al. recently reported that in the rhombohedral-I ($R$-I) phase (≈12–39 GPa) a dome-shaped SC $T_c$ peaking near ~17 K at ~30 GPa develops, advancing the view that oxygen-lattice vibrations in nearly close-packed O-layers supply a light-element-dominated EPC, an oxide-based analogue of hydride SC [11]. This perspective resonates with the design logic of superhydrides and other light-element SCs and potentially makes $ReO_3$ a conceptual bridge between metallic oxides and hydride-like phonon mediated SC [1,11].

Several questions nevertheless remain regarding the superconducting (SC) and structural evolution of $ReO_3$ under high pressure. First, the pressure range over which SC is of the bulk nature, rather than filamentary or percolative, has not been fully clarified. Second, the microscopic origin of the higher-pressure structural reconstruction, and its connection to the suppression of SC remain to be further understood. In particular, whether the high-pressure state corresponds primarily to a symmetry-lowering distortion of the rhombohedral phase or to a more substantial oxygen-framework reconstruction involving changes in Re-O coordination remains under active investigation.

In this work, we present a comprehensive, high-pressure investigation of $ReO_3$ that establishes a unified relation between structural evolution, electron–phonon coupling, and SC framework across an extended pressure-temperature space. By combining synchrotron powder X-ray diffraction, Raman spectroscopy, electrical transport, dc magnetization measurements, and first-principles calculations, we map out a sequence of pressure-induced structural transitions and correlate them directly with phonon behavior, electronic structural responses, and SC properties. We demonstrate that bulk superconductivity is stabilized within the rhombohedral phase, where cooperative octahedral rotations are correlated with enhanced EPC involving both Re- and O-associated phonon modes. First-principles calculations reveal strong electron-phonon coupling in the $R\bar{3}c$ phase, with substantial contributions from both low-frequency Re-associated modes and high-frequency oxygen-related modes, yielding a calculated SC transition temperature comparable to experimental results. Upon further compression, a symmetry-lowering structural reconstruction suppresses the bulk SC. Whereas the experimental diffraction patterns can be best described by an $R32$-like average distortion, enthalpy calculation instead predicts a lower-symmetry $mP16$-$P2/c$ phase driven by phonon instability of the rhombohedral state. This reconstructed higher-coordination phase has a reduced density of states at the Fermi level, weaker EPC, and a much lower calculated $T_c$, providing a microscopic explanation for the loss of bulk SC in the higher-pressure phase. In addition, we uncover robust NTE across multiple pressure-stabilized phases and identify a curious low-temperature volume anomaly below 35 K that is probably related to the SC regime, potentially signaling enhanced lattice instability in proximity to the high-pressure symmetry-breaking transition. Together, these results highlight the central role of lattice symmetry, oxygen-framework reconstruction, and EPC in governing SC in compressed $ReO_3$ and demonstrate pressure as a clean tuning parameter for engineering emergent quantum states in simple 5$d$ oxides.

## Results and discussion

The compiled, pressure-driven phase diagram is summarized in Fig. 1, demonstrating the electronic, magnetic, structural, and phonon vibrational properties of $ReO_3$. The vertical dashed lines in Fig. 1 mark the phase boundaries identified primarily from powder X-ray diffraction (PXRD) measurements. These boundaries are consistent with the transport, magnetization, and Raman data. Whereas minor offsets in the critical pressures are unavoidable due to the use of different pressure-transmitting media (PTMs), manometers, criteria, and measurement temperatures, the overall interpretation of the phase diagram remains robust. Five distinct pressure regimes can be identified, spanning the low-pressure $Pm\bar{3}m$ cubic phase, an intermediate pressure distorted $Im\bar{3}$ cubic phase and two-phase coexistence region ($Im\bar{3}$ cubic and $R\bar{3}c$ rhombohedral), the fully developed rhombohedral $R\bar{3}c$ phase, and a higher-pressure phase proposed on the basis of both experimental observations and theoretical analysis. These regimes are corroborated by clearly defined discontinuities in the RRR, changes in the resistance isotherms (e.g. 250 K), rotation angle of the $ReO_3$ octahedra and the emergence, splitting, and suppression of Raman-active phonon modes.

Viewed from a microscopic perspective, the phase diagram shown in Fig. 1 is most naturally organized by the progressive rotation and distortion of corner-sharing $ReO_6$ octahedra. Increasing pressure incrementally reconstructs the oxygen sublattice and its coupling to the electronic degrees of freedom, with the octahedral rotation angle acting as a measure of how the lattice responds to pressure. In the cubic phases ($Pm\bar{3}m$ to $Im\bar{3}$), $ReO_3$ remains non-SC. Cooperative octahedral rotation gradually stabilizes the rhombohedral $R\bar{3}c$ structure, in which the oxygen layers evolve toward a nearly hexagonal close-packed configuration. Bulk SC emerges exclusively within this phase and follows a dome-shaped pressure dependence, with $T_c$ increasing systematically with the rotation angle and reaching its maximum near a critical angle of ~30°, where the structural distortion is optimized. By further increasing pressure above ~ 35 GPa, our indexing suggests an effective doubling of the lattice parameters, along all crystallographic directions. Such a reconstruction of the real-space unit cell is expected to strongly modify the Brillouin-zone topology and electronic band structure through zone folding, leading to a redistribution of the electronic density of states at the Fermi level and a disruption of the phonon modes that optimally couple to the electrons. As a result, the delicate balance between lattice geometry, oxygen-dominated phonons, and EPC established in the $R\bar{3}c$ phase could be significantly altered, providing a reasonable explanation for the suppression of SC at the highest pressures.

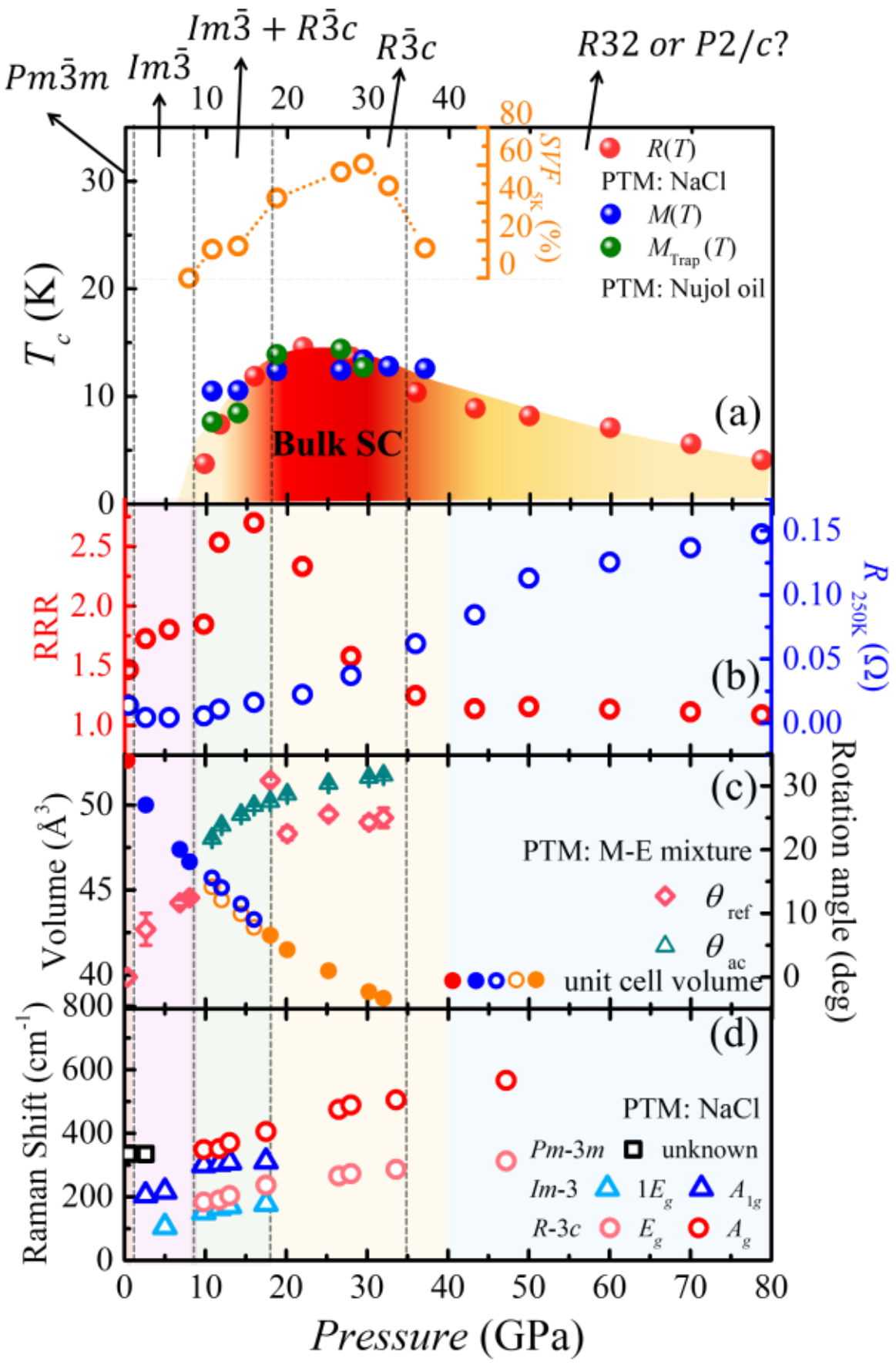

Fig. 1 **Pressure dependence of electronic, magnetic, structural, and Raman properties of $ReO_3$.** The pressure dependence of (a) superconducting transition temperature $T_c$; (b) left: residual resistance ratio, $R_{250K}/R_{20K}$, and right: resistance at 250 K; (c) left: unit cell volume estimated from the high-pressure x-ray diffraction measurement, with different colors of circles marking the evolution of the unit cell volume for five pressure regions, filled symbols stand for single -phase region ($Im\bar{3}$ or $R\bar{3}c$), open symbols for mixed-phase regions ($Im\bar{3}$ and $R\bar{3}c$); and right: rotation angle of the $ReO_6$ octahedral, in which $\theta_{ref}$ is extracted from the refined structures and $\theta_{ac} = \cos^{-1}(\sqrt{6}a/c)$ assumes rigid $ReO_6$ octahedra rotation; and (d) Raman mode frequencies. Dashed lines and different color areas mark the phase boundaries and different structures determined from PXRD measurements with the structural phases labeled above the panel (a). The uncertainty in $T_c$ arising from the temperature resolution and the determination criterion is smaller than the symbol size. The pressure values were determined at room temperature, and the pressure uncertainty is affected by pressure gradients and the protocol-dependent pressure drift upon cooling.

Figure 1(a) additionally shows a clear distinction between the SC dome inferred from transport measurements and the narrower pressure window in which bulk SC is delineated by magnetization measurements. Although SC feature persists over a broad pressure range, magnetic susceptibility measurements reveal a substantial shielding fraction only between ~20 and 31 GPa, coinciding with the pressure interval where the $R\bar{3}c$ phase is structurally dominant. Below this range, SC is suppressed by phase coexistence, while above it, increasing structural complexity and pressure gradients destabilize the bulk SC state.

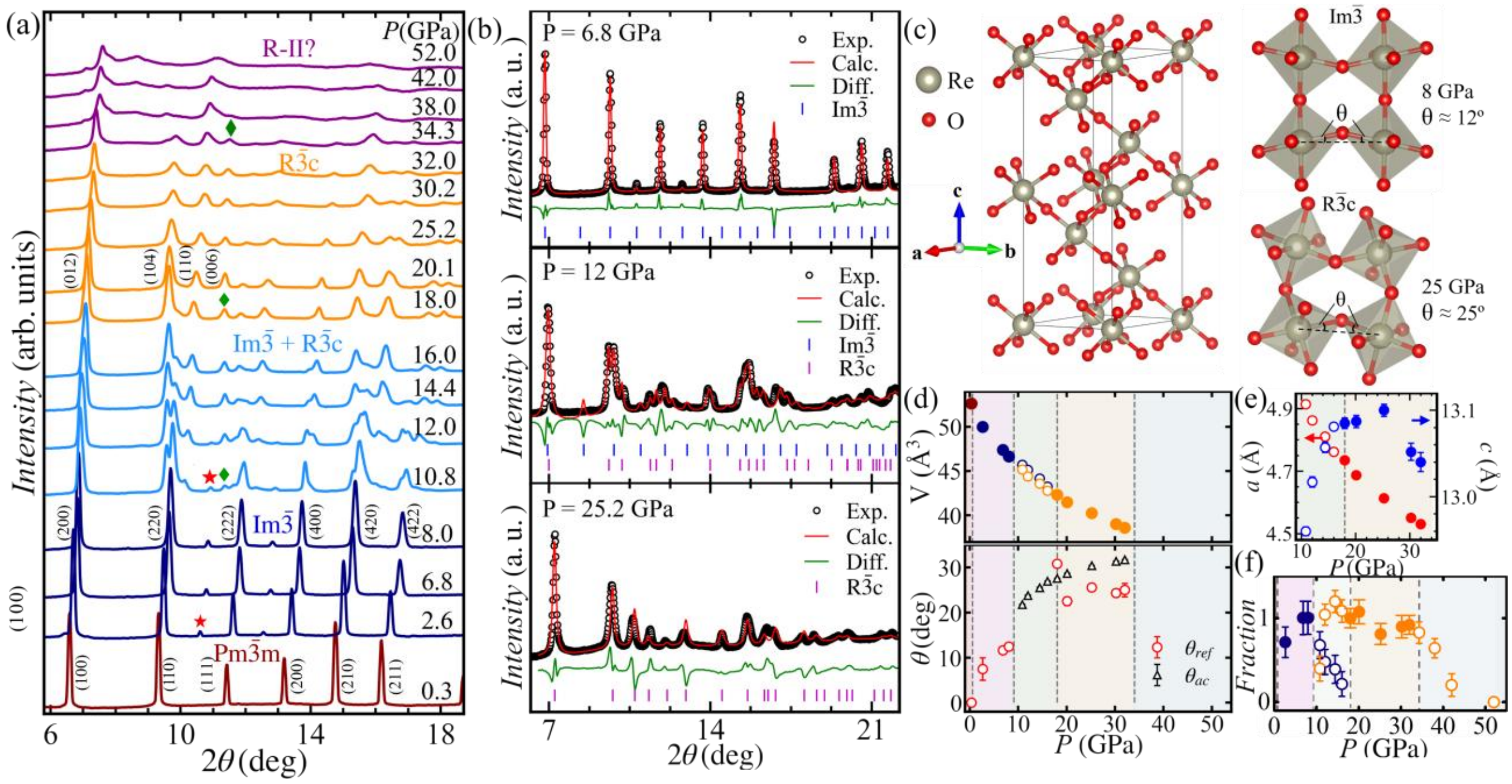

Fig. 2: **Pressure induced structural phase transitions of $ReO_3$.** (a) Room temperature powder X-ray diffraction patterns obtained on $ReO_3$ samples at various pressures from 0.3 to 52 GPa (run 1). The synchrotron X-ray wavelength is 0.4306 Å. The red star and green diamond highlight isolated peaks from $Im\bar{3}$ and $R\bar{3}c$ phases, respectively. (b) Rietveld refinement of PXRD patterns of $Im\bar{3}$ (top panel) and $R\bar{3}c$ (bottom panel) phases at 6.8 GPa and 25.2 GPa, respectively. In the middle panel, Le Bail fits two phases admixture of both $Im\bar{3}$ and $R\bar{3}c$ at 12 GPa. (c) Crystal structure of the rhombohedral $R\bar{3}c$ phase, with corner-shared $ReO_6$ octahedra. $Im\bar{3}$ and $R\bar{3}c$ phases are connected by octahedra rotations, with the rotation angle $\theta$ increasing from 12 to 25 deg between 8 GPa and 25 GPa. (d) Pressure dependence of volume per formula unit $V$ (top panel) and octahedra rotation angle $\theta$ (bottom panel). The rotation angle was determined from the refined structures $\theta_{\mathrm{ref}}$ (in red) and also estimated from lattice parameters $\theta_{ac} = \cos^{-1}(\sqrt{6}a/c)$ (in black). (e) Pressure dependences of the refined lattice constants $a$ and $c$ in the $R\bar{3}c$ phase. (f) Phase fractions $p$ in each pressure range. $p$ is estimated by the maximum intensity of the isolated peaks highlighted in (a) normalized by the corresponding intensity observed in the single-phase regions. Solid and open symbols show data points in single- or mixed-phase regions, respectively.

Figure 2(a) presents PXRD patterns obtained on $ReO_3$ for pressures ranging from 0.3 to 52 GPa. Five distinct regions are identified: i) near the ambient pressure, $P \leq 0.3$ GPa, $ReO_3$ exhibits cubic symmetry with $Pm\bar{3}m$ (space group (SG) # 221); ii) for $2.3 < P < 8$ GPa, weak reflections appear between the main peaks of $Pm\bar{3}m$ phase, and the patterns were indexed to cubic $Im\bar{3}$ symmetry (SG #204), see top panel of Fig. 2(b); iii) above 10 GPa, a rhombohedral phase with rhombohedral $R\bar{3}c$ symmetry (SG #161) emerges, and for pressures up to 16 GPa, both $Im\bar{3}$ and $R\bar{3}c$ phases are indexed, as shown in the middle panel of Fig. 2(b) at 12 GPa; iv) for 18 < P < 32 GPa, only the rhombohedral $R\bar{3}c$ phase is observed, as illustrated by the refinement of the 25 GPa data in the bottom panel of Fig. 2(b); and v) at high pressures, above 34 GPa, new peaks appear at $2\theta \approx 8$ deg, while high angle peaks disappear, suggesting the emergence of a new phase, that we are not able to unambiguously identify. Table S1 provides crystallographic information of the $R\bar{3}c$ phase at 25.1 GPa.

It is noteworthy that up to 10 GPa, our data could be readily indexed by $Pm\bar{3}m$ or $Im\bar{3}$ cubic phases, with no evidence of monoclinic $C2/c$ [12,13] or tetragonal $P4/mbm$ symmetries [14], in agreement with recent works on $ReO_3$ [11]. Recently, Efthimiopoulos et al. also suggested that the $Im\bar{3}$ phase is stable at even higher pressures, P ≈ 15 GPa, attributing the emergence of new peaks above 10 GPa to radiation damages of samples in a high-flux synchrotron source [4]. Our data is not consistent with this scenario, in fact the main reflections of $Im\bar{3}$ phase decrease in intensity above ~10 GPa and completely vanish above 16 GPa, with the $R\bar{3}c$ phase being stable up to 32 GPa. Furthermore, no anomalous peak broadening was found above 10 GPa, as would be expected for damaged samples that were still illuminated by the X-ray beam.

The pressure evolution of the $ReO_6$ framework can be described more quantitatively by the octahedral rotation angle $\theta$, as shown in Fig. 2(d). Here, $\theta$ was evaluated in two ways: $\theta_{ref}$ determined from the refined structural model and $\theta_{lat}$ estimated from the pressure-dependent lattice parameters under a rigid-octahedron approximation. $\theta_{ref}$ and $\theta_{lat}$ provide consistent evidence for progressive rotation of the corner-sharing $ReO_6$ framework upon compression. In the low-pressure cubic phases, $\theta$ is small. Upon entering the rhombohedral phase, $\theta$ increases rapidly, reflecting cooperative tilting of the $ReO_6$ octahedra. In the pressure range where bulk superconductivity is established by both electrical transport and magnetic susceptibility measurements, $\theta$ already approaches the critical value of ~30°. The lattice-parameter-derived $\theta_{lat}$ shows a weak continued increase with pressure within this bulk-superconducting region, whereas $\theta_{ref}$ obtained directly from the refined structures remains nearly pressure independent, staying close to ~30° within the uncertainty of the refinement. This behavior suggests that the rhombohedral $ReO_6$ framework has approached a geometrically optimized/saturated, strongly rotated configuration, rather than undergoing a large additional internal distortion of individual $ReO_6$ octahedra.

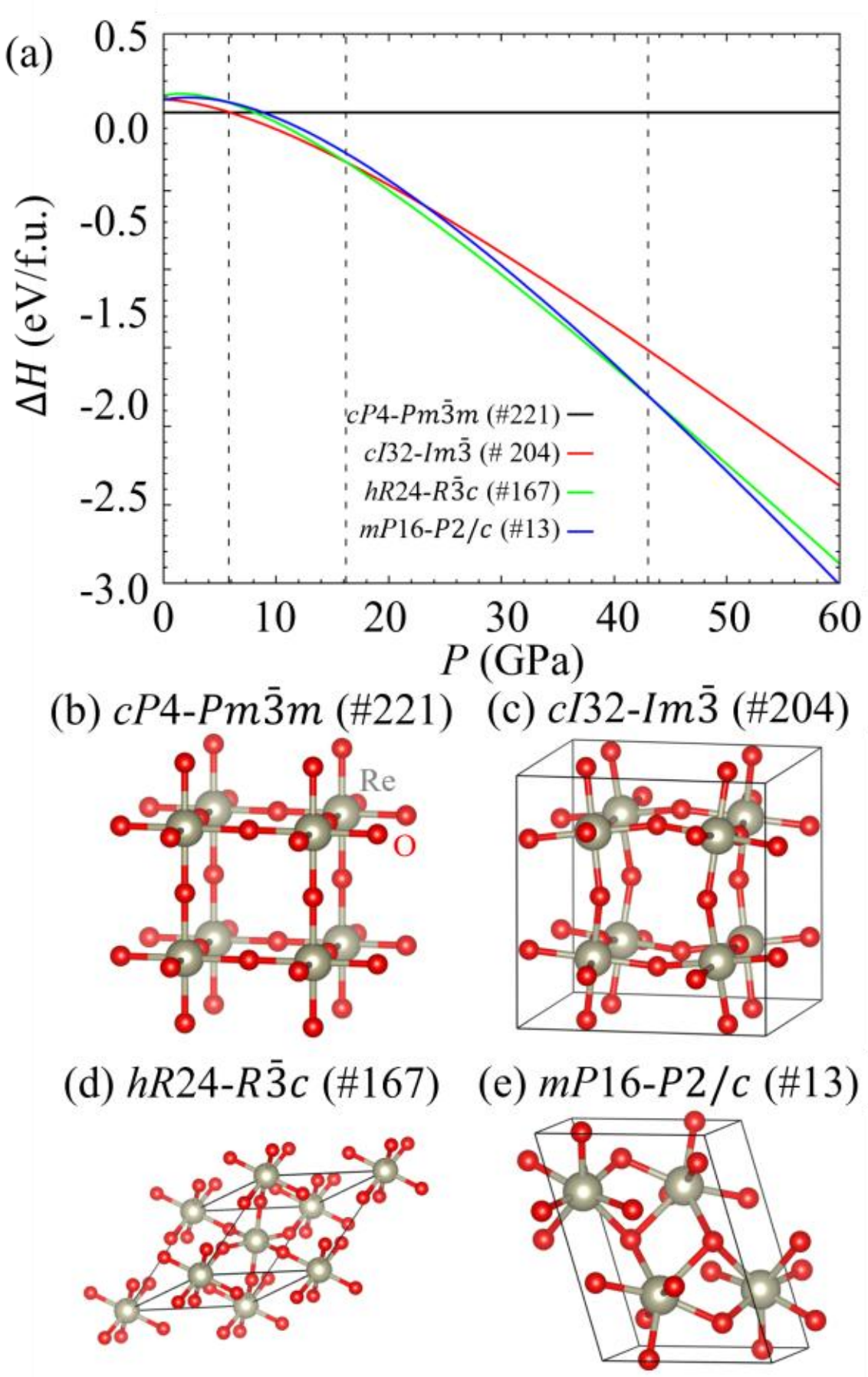

Fig.3 **Calculated structural phase transitions of $ReO_3$ in DFT.** (a) Enthalpy difference $\Delta H$ as a function of pressure ($P$) with respect to the ambient phase of $cP4\text{-}Pm\bar{3}m$ in space group 221. The calculated structural phase transitions are $cI32\text{-}Im\bar{3}$ (#204) at 5.8 GPa, $hR24\text{-}R\bar{3}c$ (#167) at 18.2 GPa and *mP16-P2/c* (#13) at 43.0 GPa. The corresponding crystal structures are displayed in (b) $cP4\text{-}Pm\bar{3}m$ (#221), (c) $cI32\text{-}Im\bar{3}$ (#204), (d) $hR24\text{-}R\bar{3}c$ (#167) and (e) *mP16-P2/c* (#13).

The sequence of the structural transitions $Pm\bar{3}m \rightarrow Im\bar{3} \rightarrow R\bar{3}c$ is consistent with the reported in ref. [11]. In that work, the authors suggested that the high-pressure phase, P > 34 GPa, also has rhombohedral symmetry (indicated by $R$-II). Within the resolution of our diffraction data in the first run, however, the patterns collected in this pressure range could not be unambiguously indexed. A second experimental run employing neon as the PTM yields significantly improved diffraction quality at high pressures (Fig. s1). For example, the diffraction pattern at 55 GPa can be indexed with several candidate symmetries (Figs. s3 and s4) among which an $R32$ structure (*SG* #155) dominated by the displacement modes $L1$ of $R\bar{3}c$ (notation from ISODISTORT [15,16]) provide the best agreement. This structure is associated with a doubling of the lattice parameters along all crystallographic directions.

Figure 3(a) plots the density functional theory (DFT)-calculated enthalpy difference $\Delta H(P)$ of the different $ReO_3$ crystal structures with respect to the ambient phase of $cP4$-$Pm\bar{3}m$ in space group #221. The calculated structural phase transitions are $cI32$-$Im\bar{3}$ (#204) at 5.8 GPa and $hR24$-$R\bar{3}c$ (#167) at 18.2 GPa, which agree reasonably well with the experimental observations. The structural change involves the bending of the Re-O-Re linear bond with increasing pressure. With even higher pressure, we predict the $mP16$-$P2/c$ (#13) structure at 43.0 GPa by following the imaginary phonon mode of $hR24$-$R\bar{3}c$ at high pressure. The distinct structural feature of $mP16$-$P2/c$ is the increased Re-O coordination number to 7 from 6 of the ideal and distorted $ReO_6$ octahedron units in the lower pressure structures as shown in Fig. 3(d,e). Consistent with this theoretical prediction, the 55 GPa diffraction pattern from run 2 can also be reasonably fitted by the $mP16$-$P2/c$ phase, as shown in Fig. s4. The increase in local Re-O coordination reflects a pressure-driven reconstruction toward a denser local environment, a structural tendency broadly reminiscent of high-pressure superconducting metal hydrides. However, in $ReO_3$, this denser higher-coordination phase does not enhance superconductivity; instead, superconductivity is suppressed, motivating the following comparison of the electronic structure, phonon spectrum, and EPC between $hR24$-$R\bar{3}c$ phase and the $mP16$-$P2/c$ phase.

We thus calculated phonon and EPC using density functional perturbative theory (DFPT), and EPC spectra using the isotropic Eliashberg approximation with the SC $T_c$ estimated in McMillian-Allen-Dyne formula. Fig. 4(a) plots the phonon spectra of the SC $hR24$-$R\bar{3}c$ structure at 40 GPa with atomic projections, which shows the low frequency modes < 7 THz are predominantly associated with Re motion, whereas predominantly O-associated modes are the higher-frequency modes. In Fig. 4(b), the same phonon spectra are decorated with mode-resolved EPC $\lambda_{qv}(\omega)$ projection (green dots), which shows large EPC on the low frequency modes. This behavior can also be seen in the Eliashberg spectral function $\alpha^2F(\omega)$ and its integration (dashed line) in Fig. 4(c). The contribution to the overall EPC $\lambda$ is already at 0.75 for Re-dominated low frequency modes < 7 THz, but the O-dominated modes are also important to contribute 0.47 to $\lambda$ at high frequency. Future high-pressure oxygen-isotope-effect measurements would provide a valuable direct test of the contribution of oxygen-associated phonons to superconductivity in $ReO_3$. With the calculated $\omega_{log}$ = 308.2 K and using the Coulomb potential of 0.16, the estimated $T_c$ = 20.1 K for $hR24$-$R\bar{3}c$ at 40 GPa is in a reasonable agreement with experiment.

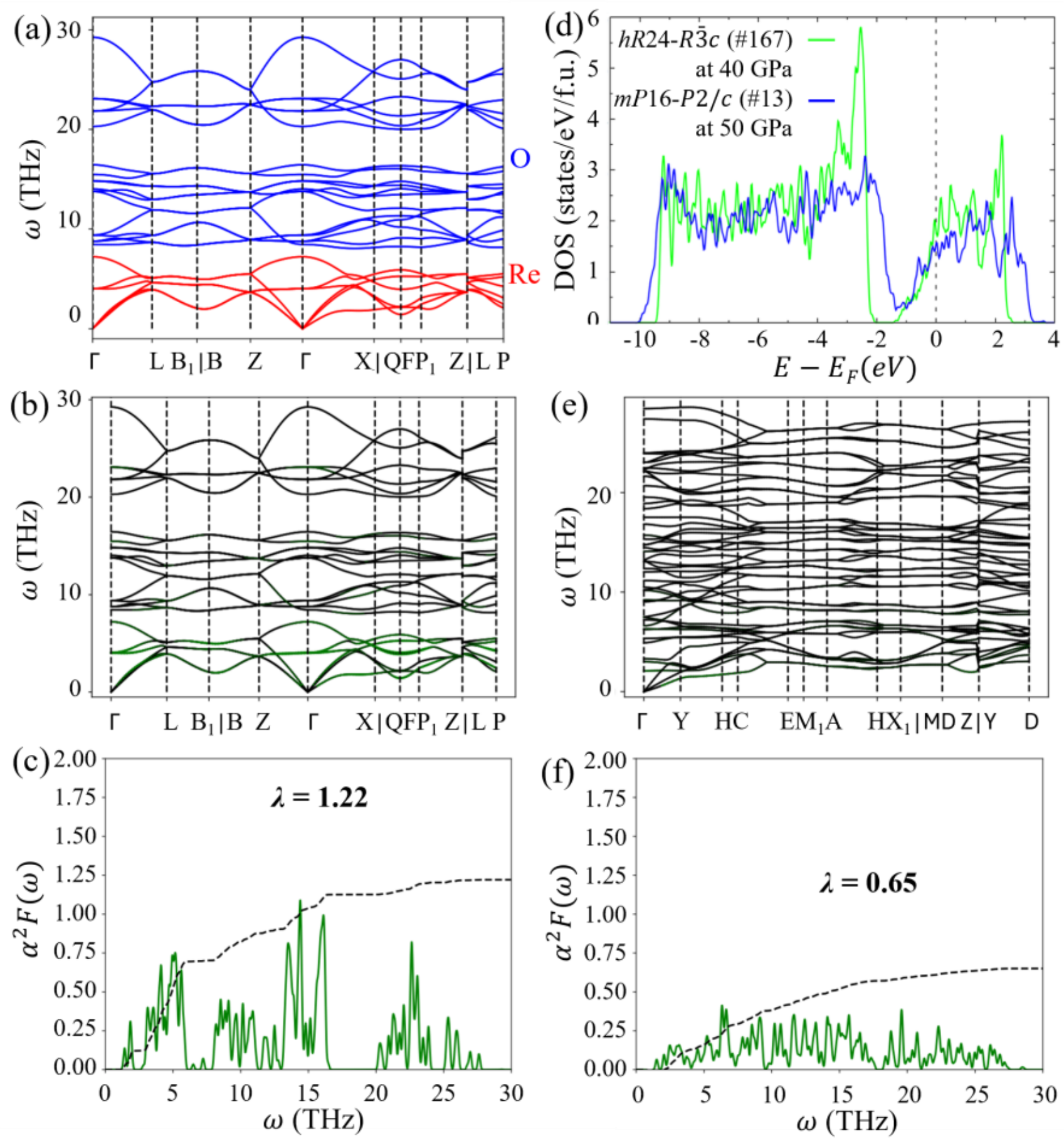

Fig. 4 **Calculated phonon and electron-phonon coupling (EPC) spectra of $ReO_3$ in *hR*24 (167) at 40 GPa and *mP*16 (13) at 50 GPa.** (a) Phonon spectra with atomic projection, (b) phonon spectra with mode-resolved EPC $\lambda_{qv}(\omega)$ projection (green dots) and (c) isotropic Eliashberg spectral function $\alpha^2F(\omega)$ and its integration (dashed line) in $hR24$-$R\bar{3}c$ (#167) at 40 GPa. (d) Electronic density of states (DOS) comparison of $mP16$-$P2/c$ (#13) at 50 GPa vs. $hR24$-$R\bar{3}c$ (#167) at 40 GPa. (e) Phonon spectra with $\lambda_{qv}(\omega)$ projection and (f) $\alpha^2F(\omega)$ of $mP16$-$P2/c$ (#13) at 50 GPa. The estimated phonon-mediated superconducting critical temperature ($T_c$) of $ReO_3$ is 20.1 K with an overall EPC $\lambda$=1.22 for $hR24$-$R\bar{3}c$ (#167) at 40 GPa and 5.8 K with $\lambda$=0.65 for $mP16$-$P2/c$ (#13) at 50 GPa.

To examine the pressure evolution of EPC within the hR24/ $R\bar{3}c$ phase, we further calculated $\lambda$, $\omega_{log}$, $T_c$, and the Re-O-Re bond angle at selected pressures (Fig. s5). With increasing pressure from 20 to 40 GPa, $\lambda$ increases from approximately 0.61 to 1.22, while the Re-O-Re angle decreases from approximately 139° to 131°, establishing a quantitative correlation between enhanced octahedral rotation and stronger EPC. However, $T_c$ does not scale monotonically with either $\lambda$ or the bond angle alone. The increase in $\lambda$ is accompanied by a decrease in $\omega_{log}$, and the competition between these two quantities produces the calculated broad dome-like $T_c$ evolution.

Now we compare the electronic structure, phonon spectrum, and EPC of the superconducting $hR24\text{-}R\bar{3}c$ phase and the higher-pressure $mP16\text{-}P2/c$ phase. As shown in Fig. 4(d), the electronic density of states (DOS), $mP16\text{-}P2/c$ at 50 GPa has a wider DOS due to shorter Re-O bonds at higher pressure than $hR24\text{-}R\bar{3}c$ at 40 GPa. Most noticeable is the decrease of the DOS at the Fermi energy ($E_F$), which also closes the pseudo gap region around $E_F$ - 2 eV. For the phonon spectra in Fig. 4(e), there are smaller mode-resolved EPC $\lambda_{qv}(\omega)$ projection than the $hR24\text{-}R\bar{3}c$ in Fig. 4(b), although with the greater number of modes due to lower symmetry. The overall EPC $\lambda$ is reduced to 0.65 as plotted in Fig. 4(f). The estimated $T_c$ is dropped to 5.8 K for the $mP16\text{-}P2/c$ at 50 GPa. This behavior of decreasing $T_c$ across the structural phase transition from $hR24\text{-}R\bar{3}c$ to a lower symmetry structure at higher pressure agrees well with experimental observation and can be explained with the structural change of $mP16\text{-}P2/c$ with the increased Re-O coordination number reducing both the DOS at $E_F$ and the overall EPC strength.

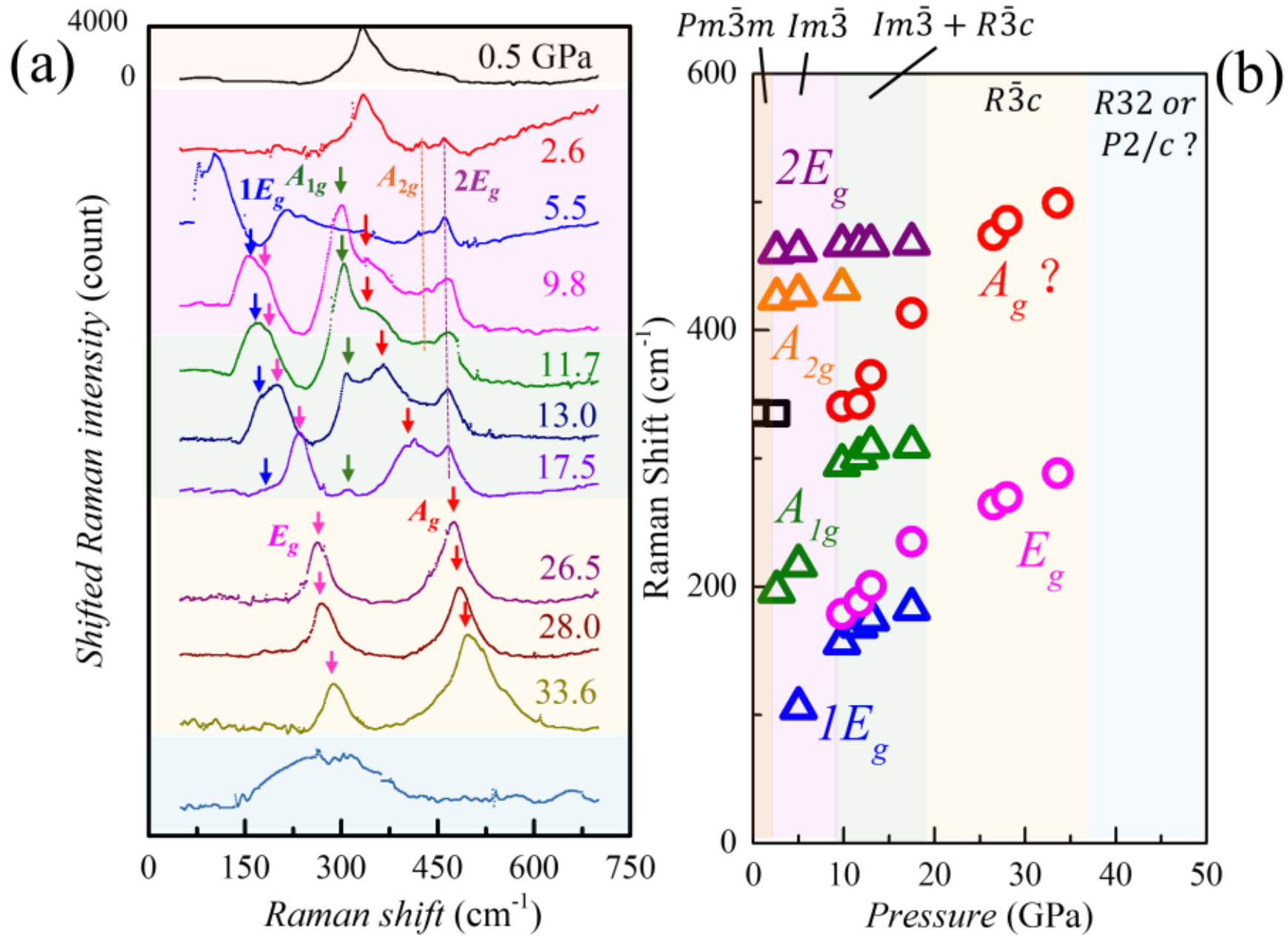

FIG. 5 (a) Room-temperature Raman spectra of $ReO_3$ at selected pressures (λ = 633 nm), vertically offset for clarity. The optical micrographs and Raman measurements were performed in the same run as the transport experiments. Colored arrows and dashed lines track the pressure evolution of symmetry-allowed phonon modes. In the $Im\bar{3}$ phase, blue and green arrows indicate the low-frequency modes ($1E_g$ and $A_{1g}$), while orange and purple dashed lines mark higher-frequency modes ($A_{2g}$ and $2E_g$, at ~ 450 $cm^{-1}$). Pink and red arrows indicate emergence of the corresponding phonon modes (possibly $E_g$ and $A_g$) in $R\bar{3}c$ phase and their pressure evolution. (b) Pressure dependence of the Raman mode frequencies extracted from (a). Symbol colors correspond to the arrows in panel (a). The black open squares denote the broad feature centered near ~340 cm$^{-1}$ observed at low pressure in $Pm\bar{3}m$ phase.

Figure 5 presents the Raman spectra at selected pressures, and the pressure evolution of the Raman mode frequencies. The optical micrographs collected at the corresponding pressures are shown in Fig. s6. Here, we emphasize that linewidths in DAC micro-Raman spectra on powders are systematically broadened by orientational averaging, grain-to-grain stress/strain distributions, and finite pressure gradients within the probed spot; therefore, we focus on the emergence/suppression and pressure evolution of dominant spectral features rather than peak sharpness. Importantly, the pressure-induced evolution of first-order Raman-active phonon modes in $ReO_3$ agrees well with the multi-step structural phase transitions shown by PXRD results. Whereas minor differences in the critical pressures may arise from the use of different PTMs, the overall Raman behavior remains qualitatively consistent with the diffraction results.

At 0.5 GPa, ReO3 in its cubic $Pm\bar{3}m$ structure exhibits an overall reddish optical contrast, as shown in Fig. s6 (the apparent color difference may be affected by the camera response, white balance, illumination conditions and the DAC optical path). The optical image, however, reveals localized grayish regions across the chamber. This non-uniform color originates from the granular nature of the loosely packed powder at very low pressure. Incomplete compaction leads to interparticle voids and local variations in optical path length, causing spatial differences in light scattering and reflectivity. Therefore, color variations at 0.5 GPa most likely reflect powder packing effects rather than the chemical inhomogeneity.

According to group-theoretical analysis for the cubic $Pm\bar{3}m$ symmetry, no first-order Raman-active phonons are allowed at the $\Gamma$ point [17]. Consistent with this expectation, the Raman spectrum at 0.5 GPa does not exhibit well-defined symmetry-resolved phonon modes. Instead, a broad feature centered near ~340 cm$^{-1}$ is observed, as shown in Fig. 5(a). This broad feature is more likely attributed to disorder-induced Raman scattering, where local structural distortions partially relax the symmetry selection rules in the nominally cubic phase. [4]. In addition, the surface oxidation in powdered samples could be another factor that leads to the broad peak [18].

Upon increasing pressure to 2.6 GPa, where the sample's structure transforms from $Pm\bar{3}m$ to $Im\bar{3}$ phase, we see much more uniform reddish color and additional low-frequency Raman modes ($A_{1g}$ and $1E_g$) become discernible, as shown in Fig. 5(a). These two modes in $Im\bar{3}$ phase have been proved to reflect the rotational/tilting motions of nearly rigid $ReO_6$ octahedra. They harden rapidly with pressure below ~5 GPa and continue to shift upward more gradually at higher pressures, reflecting progressive stiffening of the octahedral rotational degrees of freedom [4]. In our Raman spectra, the corresponding low-frequency features also shift to higher frequencies in the $Im\bar{3}$ phase, in a similar manner as shown in Fig. 5(b).

Upon further compression into the $Im\bar{3}$ / $R\bar{3}c$ coexistence region (~9.8 GPa), these modes split and reorganize, coinciding with the pressure range where our PXRD data indicate the emergence of the rhombohedral phase. Since the $R\bar{3}c$ structure is generated by cooperative rotations of the $ReO_6$ octahedra, the splitting and reorganization of these low-frequency Raman features upon entering the $Im\bar{3}$/ $R\bar{3}c$ coexistence region are consistent with a symmetry-lowering reconstruction of the rotated $ReO_6$ framework. It is noteworthy that this peak splitting was also reported by I. Efthimiopoulos et al. and was attributed to a possible solidification of PTM (4:1 methanol-ethanol mixture) [4]. We used NaCl as PTM for Raman measurements, which is not expected to have abrupt changes near ~10 GPa, so we attribute the observed splitting primarily to the intrinsic structural evolution and/or phase coexistence.

To further assess the lattice response associated with this structural evolution, we further analyzed the pressure-dependent unit-cell volume using Birch–Murnaghan equation-of-state fits for the single-phase $Im\bar{3}$ and $R\bar{3}c$ regions (Fig. s2). Both PXRD runs yield a smaller fitted bulk modulus $B_0$ for the higher-pressure $R\bar{3}c$ phase than for the lower-pressure $Im\bar{3}$ phase, suggesting enhanced compressibility of the rhombohedral phase. Based on these phase-specific bulk moduli and the pressure-dependent Raman mode frequency, a semi-quantitative mode Grüneisen-parameter analysis further suggests enhanced compression sensitivity of the low-frequency $A_g$ and $E_g$ modes in the $R\bar{3}c$ phase, as discussed in the Supplementary Information.

Above 9.8 GPa, the optical contrast of the sample darkens progressively with increasing pressure and appears nearly black by ~28 GPa. Although a faint residual reddish contrast may still be discerned at ~26.5 GPa, the Raman spectra at this pressure already display two dominant phonon modes consistent with the rhombohedral $R\bar{3}c$ framework. Group-theoretical analysis for $R\bar{3}c$ symmetry allows four Raman-active modes ($A_g$ + $3E_g$). In many $R3c/R\bar{3}c$ -type oxides, the $A_g$ mode is typically the most intense, whereas the $E_g$ modes are weaker and often partially overlapping. Consistent with this intensity hierarchy, the stronger band observed here is plausibly assigned to the $1A_g$ mode, while the lower-intensity band on its lower-frequency side can be attributed to one of the $E_g$ components. The remaining symmetry-allowed $E_g$ modes may not be individually resolved in our loading. Analogously to distorted variant of cubic $ReO_3$ structure: $TeO_3$ [19], one $E_g$ mode may lie close in frequency to $1A_g$ and merge into a single broadened feature due to systematic linewidth broadening. Another $E_g$ mode may shift toward higher frequencies with compression and extend beyond our current spectral window (~700 $cm^{-1}$), rendering it experimentally inaccessible.

Upon further compression to 47.1 GPa, where PXRD results suggests the possible emergence of a high-pressure $R32$ or $P2/c$ phase, the Raman response changes qualitatively: the higher-frequency $1A$-related feature becomes strongly suppressed, replaced by two discernible weak features, located near ~ 600 $cm^{-1}$, whereas the lower-frequency region evolves into a broadened bump with possible multi-subcomponents, as shown in Fig. 5(a). Given the absence of inversion symmetry in the $R32$ or $P2/c$ structure and the systematic linewidth broadening inherent to DAC powder measurements, a definitive mode assignment is not attempted here. A more detailed vibrational characterization of this phase would require measurements under improved hydrostatic conditions and, ideally, polarization-resolved Raman studies on single-crystals.

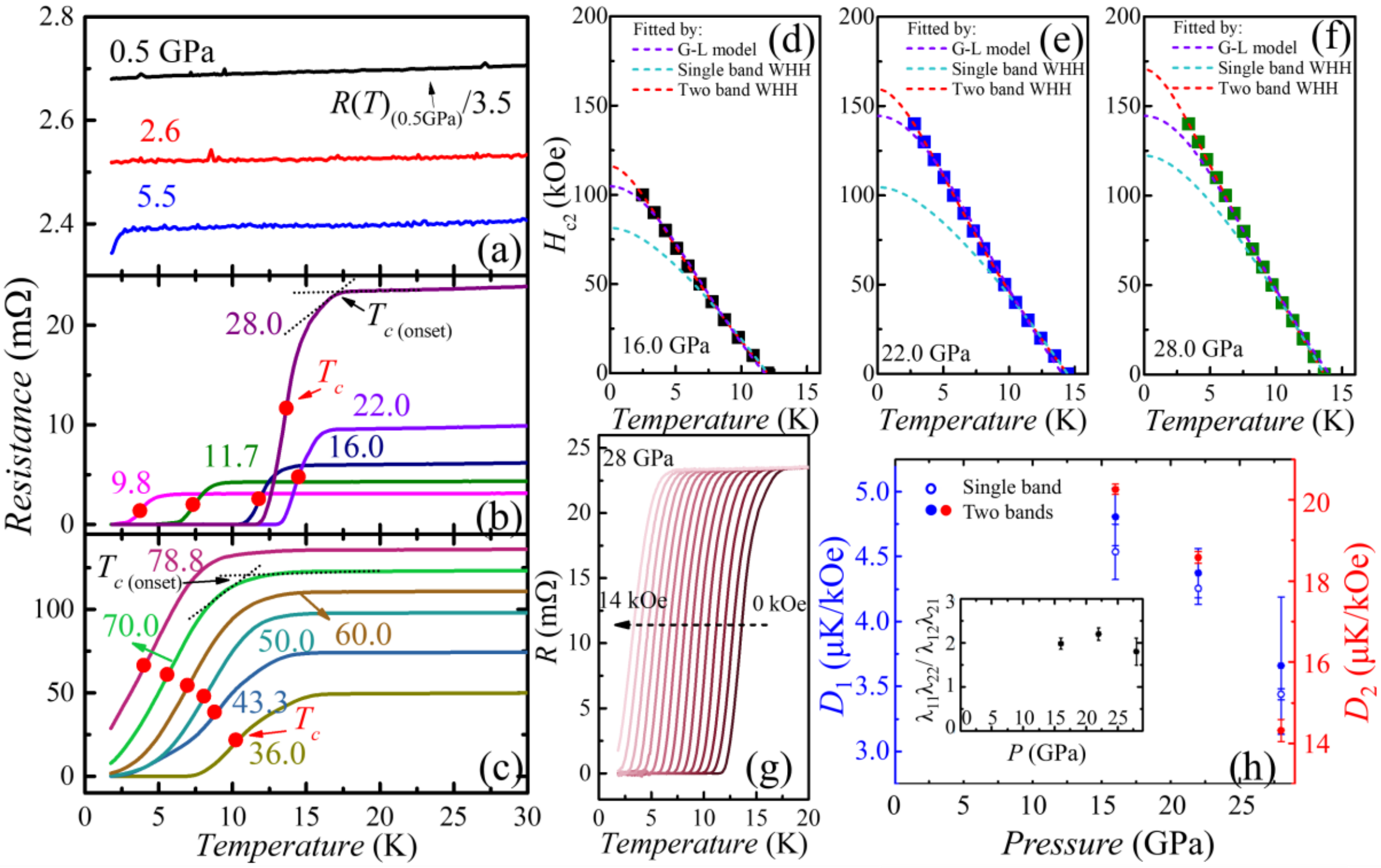

**Fig. 6 Superconductivity properties of powder $ReO_3$ under pressure** (a)-(c) Temperature dependence of the electrical resistance $R(T)$ in the low-temperature range at selected pressures from 0.5 to 78.8 GPa, illustrating the evolution of the superconducting transition under compression. The criterion for defining the superconducting transition temperature $T_c$ is the midpoint of the superconducting transition [red symbols in panels (b) and (c)]. Whereas, $T_{c,onset}$ is determined from the intersection of two extrapolated dashed lines drawn along the normal-state and transition-region slopes, as shown in panel (b). (d)-(f) Upper critical field $\mu_0 H_{c2}$ as a function of temperature at $P$ = 16.0, 22.0, and 28.0 GPa, respectively. $H_{c2}$ ($T$) values are determined from 50% resistance criterion relative to the normal-state resistance. (g) $R(T)$ at different magnetic fields at 28.0 GPa. (h) The pressure dependent fitting parameters obtained from two band WHH model.

Electrical resistance measurements under pressure, shown in Figs. 6(a-c) and Figs. s7(a-c), reveal a pronounced evolution of the normal-state transport properties of $ReO_3$. At low pressure, below 5.5 GPa, the resistance exhibits a typical metallic temperature dependence with a relatively small residual resistance, which decreases with pressure. Upon increasing pressure above approximately 5.5 GPa, the overall magnitude of resistance increases systematically over the entire measured temperature range, indicating a significant modification of the electronic transport by application of compression (as shown for 250 K in Fig. 1). This resistance enhancement persists to the highest pressures measured up to ~ 80 GPa, as shown in Figs. 1 and s7. Remarkably, a low-temperature sharp drop in resistance is observed starting at 5.5 GPa and becoming clearer above 9.8 GPa, signaling the onset of SC. The maximum $T_c$ (onset) reaches 17.5 K at 28 GPa. Alternatively, the maximum $T_c$ is 14 K at 22 GPa, as shown in Fig. 6(b), if the 50% criterion for $T_c$ is used. The normal-state resistance above the transition remains comparatively large, indicating that the emergence of SC in compressed $ReO_3$ is not associated with a reduction of the normal-state resistance. The width of the resistive signature associated with $T_c$ above 36 GPa becomes much broader and the loss of the zero resistance above 50 GPa suggests the gradual loss of the bulk nature of the SC and/or the enlargement of the pressure gradient.

To further confirm and characterize the SC phase, we measured $R(T)$ under different magnetic fields. A representative dataset recorded at 28.0 GPa is shown in Fig. 6(g). As the magnetic field is increased, the SC transition is progressively suppressed, providing further confirmation that the transition in resistance is consistent with the onset of SC. The transition width does not show substantial additional broadening under magnetic field, and similar behavior is observed at 16.0 GPa and 22.0 GPa (Fig. s7). This behavior indicates that the superconducting component contributing to the resistive transition has a relatively well-defined field scale, without a broad distribution of local $T_c$ or $H_{c2}$ values and with the irreversibility field being very close to the $H_{c2}$. However, this should not be interpreted as evidence that the entire sample is structurally homogeneous at all pressures. In particular, 16.0 GPa lies close to the $Im\bar{3}/R\bar{3}c$ coexistence regime according to the room temperature PXRD results shown in Fig. 2. In addition, the use of solid NaCl as the PTM in the transport measurements may create a different local pressure environment from that in the PXRD measurements, leading to possible differences in the nominal pressure scale and phase-coexistence range.

We also plot the temperature at which the resistance drops by 50% from the normal-state value under various magnetic fields at different pressures (16, 22, and 28 GPa) and fit the data using both Ginzburg-Landau (G-L) [20], single band [21], and double band Werthamer-Helfand-Hohenberg (WHH) [22].

The G-L model follows the expression:

$$H_{c2}(T) = \frac{H_{c2}(0)(1-t^2)}{1+t^2} \tag{1}$$

where $t = T/T_c$. The single band WHH model follows the equation:

$$H_{c2} = \frac{2\pi k_B T}{D_1 e}\left[1 - 2\psi^{-1}\left[\ln\frac{T}{T_c} - \psi\left(\frac{1}{2}\right)\right]\right] \tag{2}$$

where $\psi$ is the digamma function, and the band diffusivity $D_1 \propto \frac{\partial H_{c2}}{\partial T}|_{T_c}$ is the only fitting parameter. The two band model in which $H_{c2}$ and $T$ are related by the following equation:

$$a_0\left[\ln\frac{T}{T_c}+\psi\left(\frac{1}{2}+\frac{D_1eH_{c2}}{2\pi k_BT}\right)\right]\left[\ln\frac{T}{T_c}+\psi\left(\frac{1}{2}+\frac{D_2eH_{c2}}{2\pi k_BT}\right)\right]+(1+\lambda_-)\left[\ln\frac{T}{T_c}+\psi\left(\frac{1}{2}+\frac{D_1eH_{c2}}{2\pi k_BT}\right)\right]$$

$$+(1-\lambda_-)\left[\ln\frac{T}{T_c}+\psi\left(\frac{1}{2}+\frac{D_2eH_{c2}}{2\pi k_BT}\right)\right]=0 \quad (3)$$

Where $a_0=2\frac{\lambda_{11}\lambda_{22}-\lambda_{12}\lambda_{21}}{\sqrt{\lambda_-^2+4\lambda_{12}\lambda_{21}}}$, $\lambda_-=\lambda_{11}-\lambda_{22}$, and $\lambda_{ij}$ are the intraband ($i=j$) and interband ($i\neq j$) scattering coefficients. For the case where the intraband scattering of both bands is similar, $\lambda_-\approx 0$ and $a_0\approx\frac{\lambda_{11}\lambda_{22}}{\lambda_{12}\lambda_{21}}-1$.

The fitting curves corresponding to three models above are shown in Figs. 6(d-f). Whereas the GL model describes the $H_{c2}(T)$ data well in the vicinity of $T_c$, a clear deviation emerges at low temperatures, where the experimental data points exceed the GL extrapolation and display a slight upward curvature. On the other hand, the single-band WHH model also fails to capture the low-temperature enhancement of $H_{c2}$ observed in our data. And in fact, the deviation at low temperatures is more pronounced than that obtained from the fitting of GL model. This indicates that the breakdown of the single-band description is intrinsic. Thus, we performed the two bands WHH fitting. In this case, the data at all 3 pressures can be well fitted using only three parameters: $D_1$, $D_2$ and $a_0$. The obtained parameters are plotted in Fig. 6(h). Within the two-band WHH analysis, both band diffusivities decrease monotonically with pressure from 16 to 28 GPa, with $D_1$reduced from 4.8 to 3.8 μK/kOe and $D_2$from 20 to 14 μK/kOe. Importantly, the diffusivity ratio remains nearly constant, indicating that the pronounced two-band asymmetry persists throughout this pressure range rather than evolving toward a single-band limit. In the dirty-limit picture $D_i$ is proportional to the Fermi-surface average of $v_{F,i}^2\tau_i$. The simultaneous suppression of $D_1$ and $D_2$ implies an overall reduction of the effective quasiparticle diffusivity under compression, consistent with the resistance increase under pressure. The fact that $D_1$ remains substantially smaller than $D_2$ further suggests that the low-temperature/high-field scale of $H_{c2}$continues to be governed primarily by the slower (more diffusive-limited) band, while the faster band mainly shapes the curvature of $H_{c2}(T)$near $T_c$.

The magnetic susceptibility measurements shown in Fig. 7(a) were performed at pressures from 7.8 to 37 GPa as a function of temperature under an applied magnetic field of 0.2 kOe. A pronounced diamagnetic transition first becomes observable at 10.7 GPa, with an onset $T_c$ of ~10 K and a shielding volume fraction below 20% at 5 K. In all cases, the diamagnetic transition is not complete at 5K, so that the estimation of the shielding volume fraction at 5K is an approximate underestimate, in particular for lower $T_c$ values. Upon increasing pressure from 10.7 to 29.4 GPa, both the onset $T_c$ and the shielding volume fraction systematically increase, reaching a maximum shielding fraction of about 60% at 5 K at 29.4 GPa, indicative of the emergence of bulk SC. Further compression leads to a gradual suppression of SC, with both the onset $T_c$ and the shielding volume fraction decreasing to below 20% at 5 K at 37 GPa.

In addition to conventional magnetization measurements under finite magnetic fields, trapped-flux measurements were performed to further verify the bulk nature of SC in $ReO_3$. As shown in Fig. 7(b), a clear SC transition is observed in the trapped-flux magnetic moment, which vanishes upon warming above $T_c$. Both the trapped-flux signal amplitude and $T_c$ exhibit pressure dependences consistent with the field-dependent susceptibility data in Fig. 7(a). This behavior arises from irreversible magnetic flux trapping due to vortex pinning in the SC state, leading to a persistent magnetic signal in zero applied field that cannot be accounted for by surface/filamentary SC. In addition, the observation of a trapped-flux signal provides strong complementary evidence for bulk SC. [23,24] The irreversible trapping of flux is a direct signature of vortex pinning within the bulk SC volume, confirming the intrinsic bulk nature of the SC state in $ReO_3$ in 20-32 GPa pressure range.

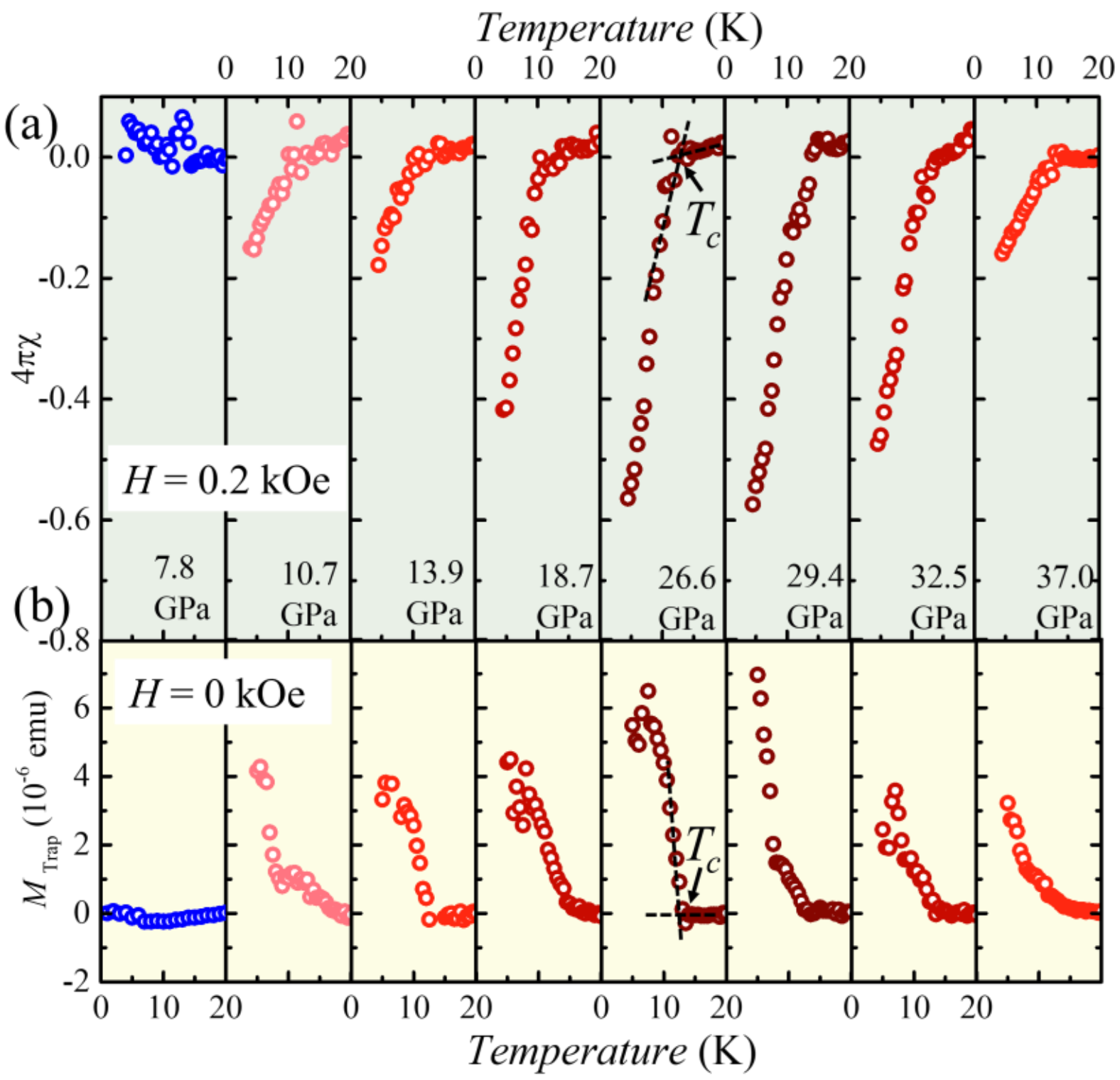

Fig. 7 **Magnetic susceptibility and trapped flux magnetization of $ReO_3$ under pressure.** (a) Zero-field-cool temperature dependent magnetic susceptibility at various pressures, with the external field $H$=0.2 kOe perpendicular to the disk-like polycrystalline sample. (b) Temperature dependent trapped flux magnetization at various pressures. The trapped flux measurements were performed by cooling the DAC from 50 to 5 K in external field, $H$ = 20 kOe perpendicular to the disk-like polycrystalline sample, stabilizing the temperature at 5 K for 10 min, and then removing the field. The magnetization was then measured upon warming from 5 to 20 K. The criteria for defining $T_c$ used here are the intersection of two extended dashed lines along the $M(T)$ and $M_{\mathrm{Trap}}(T)$ curve above and below the transition temperatures, shown in the curves of 26.6 GPa.

$ReO_3$ at ambient pressure exhibits a pronounced NTE below ~200 K, driven by thermally activated rotations (tilts) of its corner-sharing $ReO_6$ octahedra. [3] Under higher pressures, as shown in Fig. 8(a), this unusual NTE behavior persists, but its onset shifts to relatively lower temperature, and the magnitude of the lattice volume contraction is much more enhanced than at ambient pressure.

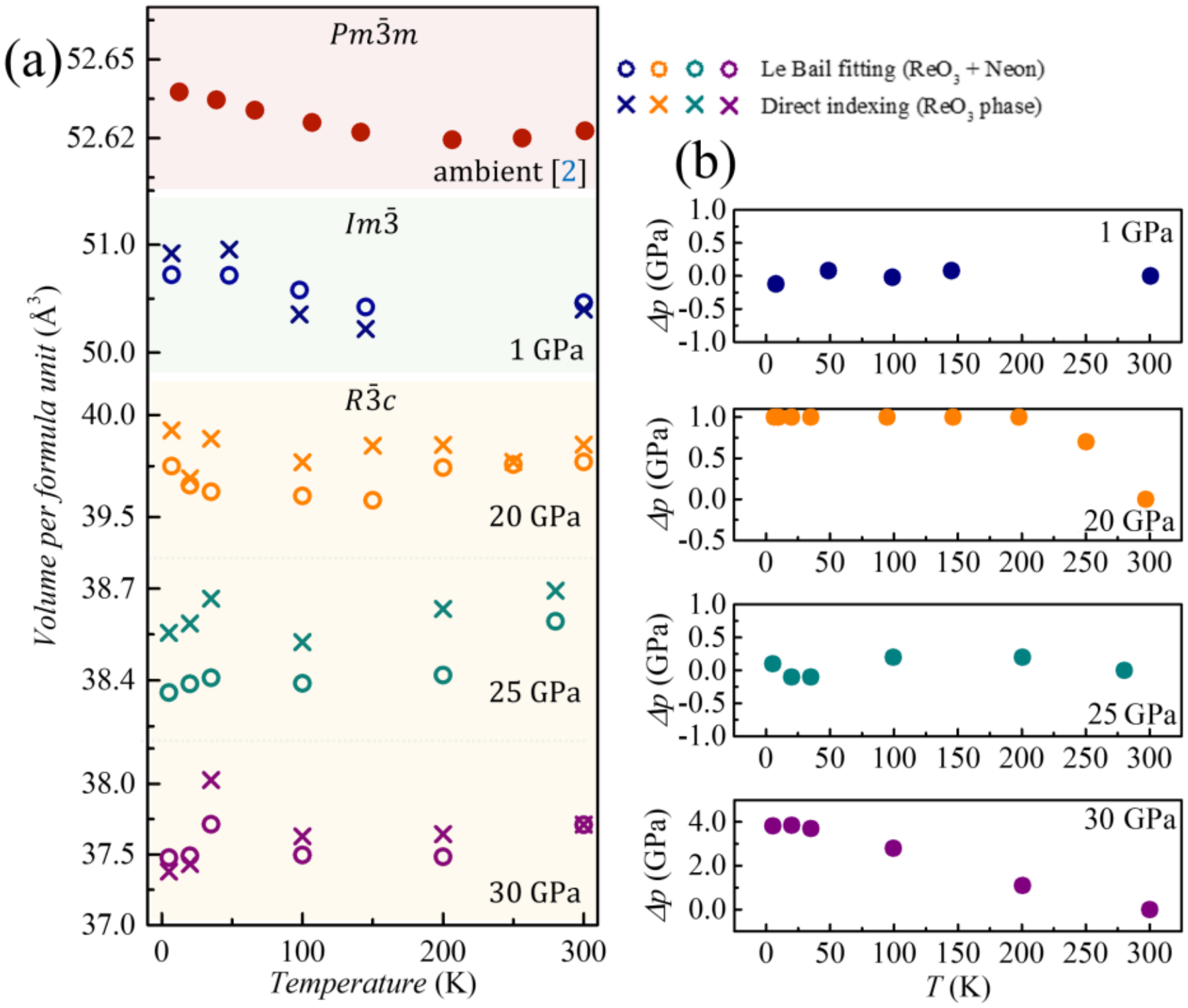

Fig. 8 **Low-temperature behavior of the unit cell volume at various pressures.** (a) Temperature dependence of the volume per formula unit at selected pressures. Red solid circles denote the ambient-pressure data digitized from Ref. [2]. Hollow circles represent lattice parameters obtained from Le Bail refinement (including $ReO_3$ and Ne phases), whereas crosses correspond to volumes extracted from direct indexing of selected reflections in the PXRD patterns (see Fig. s8 for detailed data). The weighted profile factors of the Le Bail fits are in the range $R_w$ = 3-12%, indicating reasonable refinement quality under high-pressure DAC conditions. (b) Temperature dependence of the relative pressure change, $\Delta p = p - p_{\max T}$, determined from ruby fluorescence measurements at selected pressures, allowing evaluation of possible pressure drift upon cooling.

It should be noted, though, that at high pressures the PXRD patterns (performed at beamline 3-ID-B, Advanced Photon Source, Argonne National Laboratory, which has very small spot size ~ 5 × 5 μm$^2$) exhibit pronounced azimuthal intensity inhomogeneity, manifested as spotty or segmented rings rather than uniform powder rings. This behavior primarily reflects incomplete powder averaging with much small beam size conditions in DAC, where the reduced illuminated volume limits the number of randomly oriented crystallites contributing to each reflection and enhances grain-statistics effects. Nonhydrostatic stress components due to the solidification of the PTM (Neon) at high pressure may also introduce micro-strain and partial crystallite reorientation, amplifying intensity variations along the rings. Here, we adopted two approaches to extract reliable lattice parameters. First, Le Bail fits were performed using a two-phase model ($ReO_3$ + Neon), which refines peak positions and profiles without relying on intensity constraints. Second, lattice parameters were independently determined by directly indexing selected isolated $ReO_3$ reflections, as shown in Fig. 8(a).

At 25 GPa and 30 GPa, where bulk SC is well established and the $R\bar{3}c$ phase is dominant, both lattice-parameter extraction procedures reveal a similar downturn of the unit-cell volume upon cooling below approximately 35 K. The pressure, monitored by ruby fluorescence at each temperature, remains relatively constant in this low-temperature range, as shown in Fig. 5(b), indicating that the anomaly is not simply caused by pressure drift. A comparably clear anomaly is not resolved in 20 GPa dataset. However, this absence should be interpreted with caution. The phase boundaries of $ReO_3$ depend on the PTM and experimental conditions: whereas the first room-temperature PXRD run (run 1, Fig. 2) using 4:1 methanol-ethanol identifies the $R\bar{3}c$-dominant region above approximately 20 GPa, the second PXRD run (run 2, Fig. s1) using neon shows that the $Im\bar{3}$/ $R\bar{3}c$ coexistence region can extend to nearly 26 GPa. Since the low-temperature PXRD measurements were also performed using neon as the PTM, the condition at 20 GPa may still in $Im\bar{3}$/ $R\bar{3}c$ coexistence region. Under such circumstances, local stress between coexisting phase-domains and pressure gradient could mask a subtle intrinsic lattice response of the $R\bar{3}c$ phase. Moreover, the low-temperature PXRD measurements have more limited low-angle coverage than room-temperature measurements, where several reflections diagnostic of phase coexistence are expected. Thus, the lack of a clear volume anomaly at nominal 20 GPa does not necessarily rule out its association with the more fully developed $R\bar{3}c$ superconducting regime. Meanwhile, the temperature sampling in the current dataset is relatively sparse, particularly in the 5-50 K range. Clarifying the relationship between superconductivity and this lattice anomaly will require additional high-resolution low-temperature structural studies, ideally performed under identical PTM conditions and with improved low-angle coverage, to determine the microscopic origin and physical significance of the anomaly.

## Conclusions:

In summary, our comprehensive high-pressure study of $ReO_3$ establishes a close relation between lattice symmetry, electron-phonon coupling, and the emergence and suppression of bulk superconductivity (SC). Bulk SC with the maximum onset $T_c \sim 17.5$ K emerges exclusively within the rhombohedral $R\bar{3}c$ phase, where cooperative $ReO_6$ octahedral rotations reconstruct the oxygen framework and create a favorable environment for strong electron-phonon coupling (EPC). The calculations reveal that the enhanced EPC arises from cooperative Re-O lattice dynamics, with substantial contributions from both low-frequency Re-associated modes and higher-frequency O-associated modes. Upon further compression, both experiment and theoretical calculation reveal an instability of the rhombohedral SC phase toward reconstructed lower-symmetry states associated with significant electronic reconstruction. Whereas the high-pressure diffraction patterns can be best fitted with an $R32$-like average distortion that nearly doubles the unit cell, theoretical calculations instead predict a competing $mP16$-$P2/c$ structure driven by phonon instability and increased Re-O coordination. Importantly, this denser higher-coordination phase suppresses the SC by reducing the density of states at the Fermi level and weakening the overall EPC. In addition, we uncover robust NTE across multiple pressure-stabilized phases and identify a curious low-temperature volume anomaly below 35 K that is probably related to the SC regime, potentially signaling enhanced lattice instability in proximity to the high-pressure symmetry-breaking transition. These results demonstrate that SC in compressed $ReO_3$ is governed not simply by pressure-induced densification, but by a delicate balance between oxygen-framework geometry, electronic structure, and phonon-mediated pairing in a simple $5d$ oxide.

## Methods:

### Sample acquisition and preparation

Rhenium (VI) oxide ($ReO_3$) powder (CAS No. 1314-28-9) with 99.9 % (metals basis) purity was purchased from Thermo Scientific Chemicals. To preserve sample quality, the powder container was opened and stored exclusively in an Ar-filled glovebox. Since $ReO_3$ thermally decomposes above approximately 400 °C, the small amount of powder taken for synchrotron experiments was placed in a small jar under Ar atmosphere. The jar lid was sealed with airtight membrane tape to maintain an inert environment, avoiding vacuum sealing in silica tubes that requires flame-heating and could potentially cause slight decomposition or oxygen loss during the sealing process. All experiments were conducted using the as-received powder without further purification.

### Electrical transport under high pressure, optical micrograph and low frequency Raman spectra

Electrical resistance was measured up to 80 GPa using a CuBe diamond-anvil pressure cell (DAC) (Bjscistar) compatible with a Quantum Design Physical Property Measurement System (PPMS). [25] The cell employed Type la diamond anvils with 200 µm culet size and a stainless-steel gasket. After pre-indenting the gasket to ~15 GPa, a concentric 200 µm hole was drilled by electric-discharge machining (EDM) and back-filled with a cured mixture of cubic BN powder and epoxy. This insert was then compressed to 15 GPa again and re-drilled using mechanical drilling machine to a 100 µm diameter hole to form the sample chamber.

The bottom space of the sample chamber was first covered with a thin NaCl layer along with a small ruby sphere. $ReO_3$ powder was placed above this layer, and the Pt foil electrodes, were arranged in a van der Pauw configuration on the pressed powder sample. An additional thin NaCl layer was placed over the sample and electrodes to improve the hydrostaticity at high pressure. Schematics of the sample chamber at various pressures are shown in Fig. 5a. Pressure was determined from the ruby $R1$ fluorescence line below 50 GPa and from the first-order diamond Raman edge above 50 GPa [26,27]. It is noteworthy that several pressure points between 30 and 50 GPa were calibrated using both the ruby fluorescence line and the diamond Raman edge, confirming the consistency between these two pressure gauges.

All pressures assigned to the transport data were determined at room temperature. In an auxiliary calibration (shown in Fig. s9) test using the same DAC configuration and an initial pressure of approximately 8.0 GPa an appreciable increase in the sample-chamber pressure was observed upon cooling using ruby fluorescence measurements. Because this pressure drift depends on the initial pressure and cooling protocol, no numerical correction was applied to the transport data.

Optical micrographs were recorded at each pressure using a CCD camera. For consistency, the brightness of all images was adjusted by normalizing to the intensity of the Pt wires, which serve as an internal reference. This procedure ensures that relative changes in the apparent color of the sample can be directly compared across different pressures.

Low-frequency Raman scattering measurements were performed using a Raman spectrometer (Horiba IHR-550) with a laser excitation wavelength of 633 nm. A 50 × objective lens was used to focus the incident laser beam, and all spectra at different pressures were collected from the same position near the center of the sample.

### Magnetization under high pressure

DC magnetization measurements were performed in a Quantum Design Magnetic Property Measurement System (MPMS) down to 5 K using an easyLab Mcell Ultra DAC [28] equipped with 400 μm culet anvils. An apertured tungsten gasket with a 200 μm hole served as sample chamber. Nujol mineral oil served as the pressure-transmitting medium (PTM). [29] Pressure was measured by the ruby fluorescence $R1$ line. [26,27]

Background signals of the empty DAC (the DAC was closed with all parts except for the sample) were measured at 4.4 GPa in an applied field of 0.2 kOe for magnetic susceptibility $M(T)$. The background signal of the trapped-flux magnetization measurement $M_{trap}(T)$ was also performed: the empty DAC was cooled from 30 K to 5 K in a 20 kOe external magnetic field, stabilized at 5 K for 10 min, the field was then removed, and $M_{trap\text{-}background}(T)$ was recorded while warming from 5 K to 30 K with a 0.5 K step size. The temperature was stabilized at each step for 40 seconds.

It is noteworthy that the magnetization measurements were carried out on a pressed powder sample of $ReO_3$ rather than on a bulk single crystal. Accordingly, corrections associated with the sample density and demagnetization effects were taken into account when estimating the volume shielding fraction. The pressed powder pellet was prepared in an Ar-filled glovebox using a hydraulic press with a load of 3000 lbs. The resulting pellet density was estimated to be approximately 3.67 g/cm$^3$, corresponding to about 53% of the nominal density of $ReO_3$ (6.92 g/cm$^3$), which yields a susceptibility correction factor of $f \approx 1/0.53 \approx 1.9$. A thin rectangular piece with dimensions of ~ 120 μm × 100 μm × 20 μm was cut from the pellet and loaded into the DAC, resulting in an estimated demagnetization factor of n ≈ 0.8.[30] After background acquisition, the DAC was opened, the $ReO_3$ piece was loaded into the sample chamber, and the same measurement protocols were repeated at various pressures. Sample magnetization was obtained by point-by-point subtraction of long-scan responses with and without the sample, followed by dipole fitting of the subtracted scan. [31]

### Synchrotron powder X-ray diffraction

Room-temperature high-pressure powder X-ray diffraction (PXRD) measurements were performed at beamline 13-BM-C (GSECARS), Advanced Photon Source, Argonne National Laboratory. Monochromatic X-rays with a wavelength of 0.4306 Å were focused to a 10 μm × 10 μm spot. Fresh $ReO_3$ powder, protected under an argon atmosphere, was rapidly loaded into a wide-opening BX-90 DAC [32] equipped with type Ia diamonds with 300 μm culets, a pre-indented Re gasket, and two ruby spheres (<10 μm in diameter) for pressure calibration. A methanol:ethanol (4:1) mixture was used as the PTM for run 1, while neon was employed for run 2.

Low-temperature high-pressure PXRD experiments were conducted at beamline 3-ID-B, Advanced Photon Source, Argonne National Laboratory. X-rays with a wavelength of 0.4875 Å were focused to a 5 μm × 5 μm spot. An iBX80 DAC equipped with 500 μm culet size conical anvils providing an 80° optical opening angle, was used for applying pressure. Low-temperature conditions were achieved using a Lake Shore Model RGC4 cryogen-free closed-cycle refrigeration system and a DacTools custom cryostat. Pressure was controlled in situ using a double-sided gas-membrane system. Temperature-dependent PXRD measurements at each pressure point were conducted independently. Pressure was first stabilized at 300 K (280 K for 25 GPa). The sample was subsequently cooled to the desired temperature, following by a ~20 minutes waiting time for the temperature stabilization prior to diffraction data acquisition. Pressure was monitored at each temperature point. Pressure for all measurements was determined in situ from the calibrated shift of the ruby $R1$ fluorescence line. [26,27] Two-dimensional diffraction images were integrated using DIOPTAS,[33], symmetry indexing, and Rietveld/Le Bail refinements were carried out with GSAS-II. [34]

**Computational Methods**

Electronic structure and total energy calculations in density functional theory [35,36] (DFT) have been performed with PBE [37] exchange-correlation functional using a plane-wave basis set and projector augmented wave method [38], as implemented in the Vienna Ab-initio Simulation Package [39,40] (VASP). We use a kinetic energy cutoff of 500 eV, $\Gamma$-centered Monkhorst-Pack [41] $k$-meshes with a reciprocal density of 0.025 (1/Å$^3$), and a Gaussian smearing of 0.05 eV. For enthalpy calculations of $ReO_3$ in different crystal structures, the shape of the unit cells and ionic positions are relaxed until the absolute value of force on each atom is less than 0.01 eV/Å for the volumes scaled around the equilibrium. Then these results are fit to Birch-Murnaghan [42] equation of state to evaluate pressure and enthalpy. The phonon and electron-phonon coupling (EPC) spectra are calculated in density functional perturbative theory [43] as implemented in Quantum Espresso [44] (QE) using ultrasoft pseudopotentials [45,46] with a kinetic energy cutoff of 50 Ry and a Gaussian smearing of 0.02 Ry. Spin-orbit coupling (SOC) was not included in the Quantum ESPRESSO EPC calculations. The phonon-mediated SC properties have been calculated in the isotropic Eliashberg approximation with the SC critical temperature ($T_c$) estimated in McMillian-Allen-Dyne formula [47] with a Coulomb potential of 0.16. The workflows of these first-principles calculations have been streamlined in our recently developed high-throughput electronic structure package [48-51] (HTESP).

## Acknowledgements

Work at Ames National Laboratory is supported by the US DOE, Basic Energy Sciences, Material Science and Engineering Division under contract no. DE-AC02-07CH11358. Development of capabilities for synthesis and measurements under pressure was supported by Ames National Laboratory's Laboratory Directed Research and Development (LDRD) program. RFSP's one-year visit to Ames Laboratory and Iowa State University was supported by the São Paulo Research Foundation (FAPESP), Brasil, Process Number 2024/08497-6. RFSP also acknowledges support from FAPESP under Process Number 2021/01004-6. SLM acknowledges support from FAPESP under Process Number 2023/10775-1. The computational part of work used resources of the National Energy Research Scientific Computing Center (NERSC), a DOE Office of Science User Facility. Work at Argonne National Laboratory is supported by the U.S. Department of Energy, Office of Science, under contract No. DE-AC-02-06CH11357. Portions of this work were performed at GeoSoilEnviroCARS (The University of Chicago, Sector 13), Advanced Photon Source (APS), Argonne National Laboratory. GeoSoilEnviroCARS is supported by the National Science Foundation – Earth Sciences via SEES: Synchrotron Earth and Environmental Science (EAR -2223273). This research used resources of the Advanced Photon Source, a U.S. Department of Energy (DOE) Office of Science User Facility operated for the DOE Office of Science by Argonne National Laboratory under Contract No. DE-AC02-06CH11357. Use of the COMPRES-GSECARS gas loading system was supported by COMPRES under NSF Cooperative Agreement EAR -1606856 and by GSECARS through NSF grant EAR-1634415 and DOE grant DE-FG02-94ER14466. This research used resources of the Advanced Photon Source, a U.S. Department of Energy (DOE) Office of Science User Facility operated for the DOE Office of Science by Argonne National Laboratory under Contract No. DE-AC02-06CH11357.

## Competing interests

The authors declare no competing interests.

## Data Availability

The data that support the findings of this study will be made available in DataShare, an open-access repository at Iowa State University.

## Supplemental information

# The observation of bulk superconductivity in Rhombohedral $ReO_3$ under pressure

S. Huyan [1,2*†], R. F. S. Penacchio[1,3*], L.- L. Wang[1,2*], J. Schmidt[1,2‡], D. Zhang[4], B. Lavina[4,5], Z. Li[1,2], R. A. Ribeiro[1,2], T. J. Slade[1,2], J. Zhao[5], S. L. Morelhão[3], P. C. Canfield[1,2], S. L. Bud'ko[1,2†]

[1] *Ames National Laboratory, US DOE, Iowa State University, Ames, Iowa 50011, USA*
[2] *Department of Physics and Astronomy, Iowa State University, Ames, Iowa 50011, USA*
[3] *Institute of Physics, University of São Paulo, São Paulo, SP, Brazil*
[4] *Center for Advanced Radiation Sources, The University of Chicago, Chicago, Illinois 60637, USA*
[5]*Advanced Photon Source, Argonne National Laboratory, Argonne, IL 60439, USA*

[*] These authors contributed equally.

[†] shuyan@iastate.edu, budko@ameslab.gov.

[‡] Currently at Instituto de Física de Buenos Aires, CONICET-Universidad de Buenos Aires, Pabellón 1, Ciudad Universitaria, CABA, 1428, Argentina.

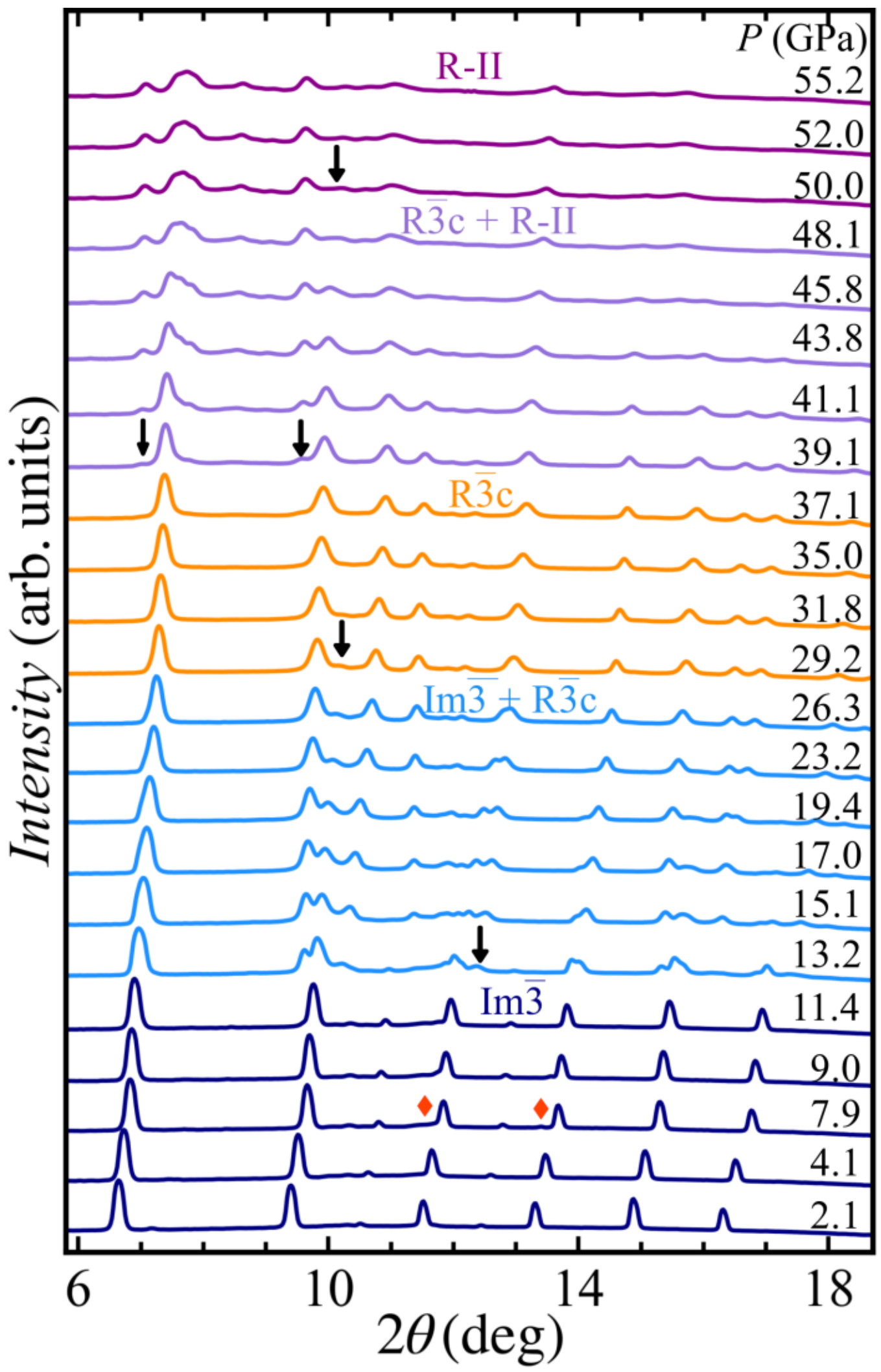

Fig. s1 High pressure powder XRD for run #2. Neon is used as pressure transmitting medium. Red diamonds denote the Neon peaks. The black arrows indicate the appearance/disappearance of main peaks that we used to define the region boundaries (one phase or two-phase).

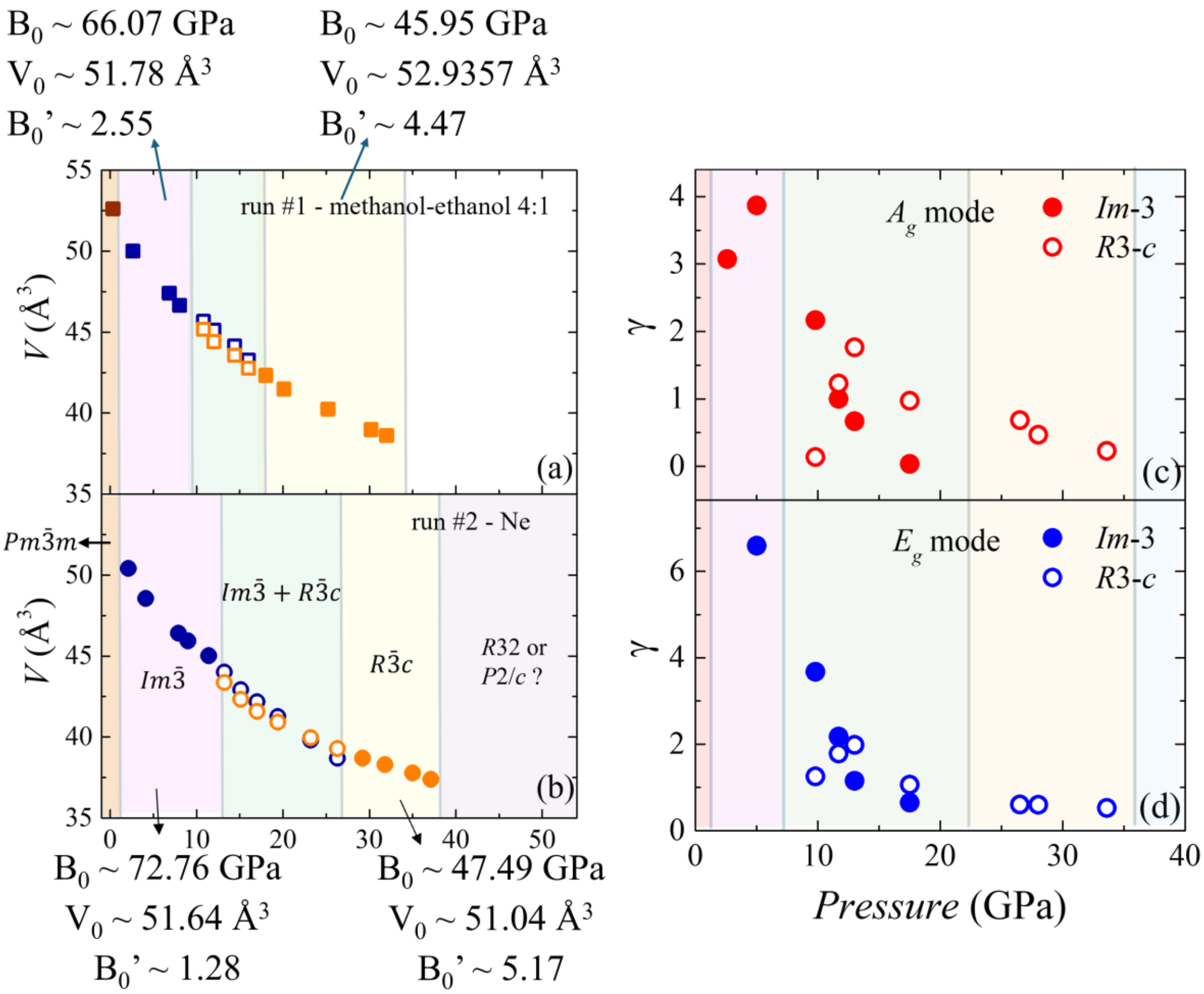

Fig. s2. Pressure dependence of the unit-cell volume and Raman-derived mode Grüneisen parameters for $ReO_3$. (a,b) Pressure-dependent unit-cell volume obtained from PXRD measurements for (a) run #1 using 4:1 methanol-ethanol as the pressure-transmitting medium (PTM) and (b) run #2 using neon as the PTM. The colored shaded regions indicate the identified structural regimes: $Pm\bar{3}m$, $Im\bar{3}$, $Im\bar{3}$ + $R\bar{3}c$ coexistence, $R\bar{3}c$, and the higher-pressure phase region. The single-phase $Im\bar{3}$and $R\bar{3}c$ regions were fitted using a third-order Birch–Murnaghan equation of state. Here, $B_0$, $V_0$, and $B_0'$ represent the zero-pressure bulk modulus, zero-pressure unit-cell volume, and the first pressure derivative of the bulk modulus, respectively. (c,d) Pressure dependence of the mode Grüneisen parameters, $\gamma$, for the low-frequency (c) $A_g$ and (d) $E_g$ Raman modes, estimated from the pressure dependence of the Raman mode frequencies using the phase-specific $B_0$ values obtained from the PXRD run #1 equation-of-state fits for the $Im\bar{3}$ and $R\bar{3}c$ phases.

The pressure-dependent unit-cell volumes provide further evidence for curious lattice response across the $Im\bar{3}$-to-$R\bar{3}c$ structural evolution. As shown in Fig. 2(a,b), the $V(P)$ data from both PXRD runs can be fitted using a third-order Birch-Murnaghan equation of state. Although the $R\bar{3}c$ phase is stabilized at higher pressure and has a smaller unit-cell volume, the fitted bulk modulus $B_0$ is consistently smaller than that of the lower-pressure $Im\bar{3}$ phase in both runs. This indicates that the high-pressure rhombohedral phase is not simply a more rigid, densely packed structure, but instead retains enhanced compressibility, likely associated with pressure-tunable soft structural degrees of freedom such as $ReO_6$ octahedral rotations/distortions. Using the phase-specific $B_0$ values together with the pressure dependence of the Raman-active low-frequency $A_g$ and $E_g$ modes shown in Fig. 5(b), we further estimated the corresponding mode Grüneisen parameters, $\gamma$. As shown in Fig. s2(c,d), $\gamma$ decreases rapidly with pressure in the $Im\bar{3}$ phase and then becomes relatively slowly pressure-dependent, whereas it has an abrupt increase upon entering the $R\bar{3}c$ phase. Although the values near the $Im\bar{3}$/ $R\bar{3}c$ coexistence regime may be affected by phase coexistence and uncertainties in assigning the Raman modes, the overall enhancement of $\gamma$ in the $R\bar{3}c$ phase demonstrates that these low-frequency modes become more sensitive to lattice compression after the rhombohedral distortion develops. This behavior may support a softened lattice response in the $R\bar{3}c$ phase and suggests that the same octahedral rotational/distortive degrees of freedom responsible for the rhombohedral distortion may also enhance the lattice contribution relevant to superconductivity in the $R\bar{3}c$ phase $ReO_3$.

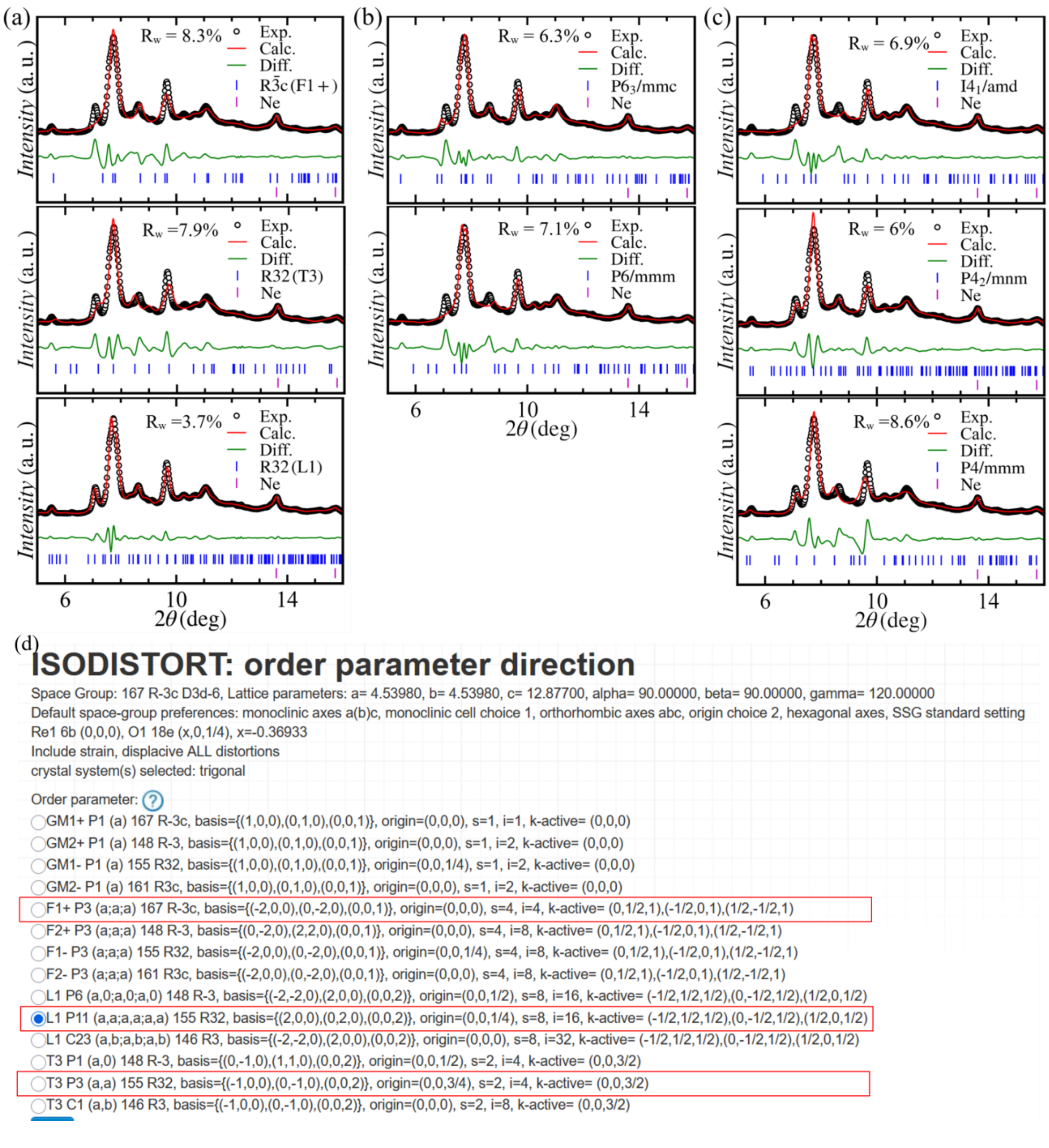

Fig. s3 Candidate indexing solutions for the 55 GPa diffraction pattern of sample #2 based on (a) rhombohedral, (b) hexagonal, and (c) tetragonal structural models. (d) Order parameters derived from ISODISTORT symmetry-mode analysis. Detailed indexing parameters are listed in Table S2.

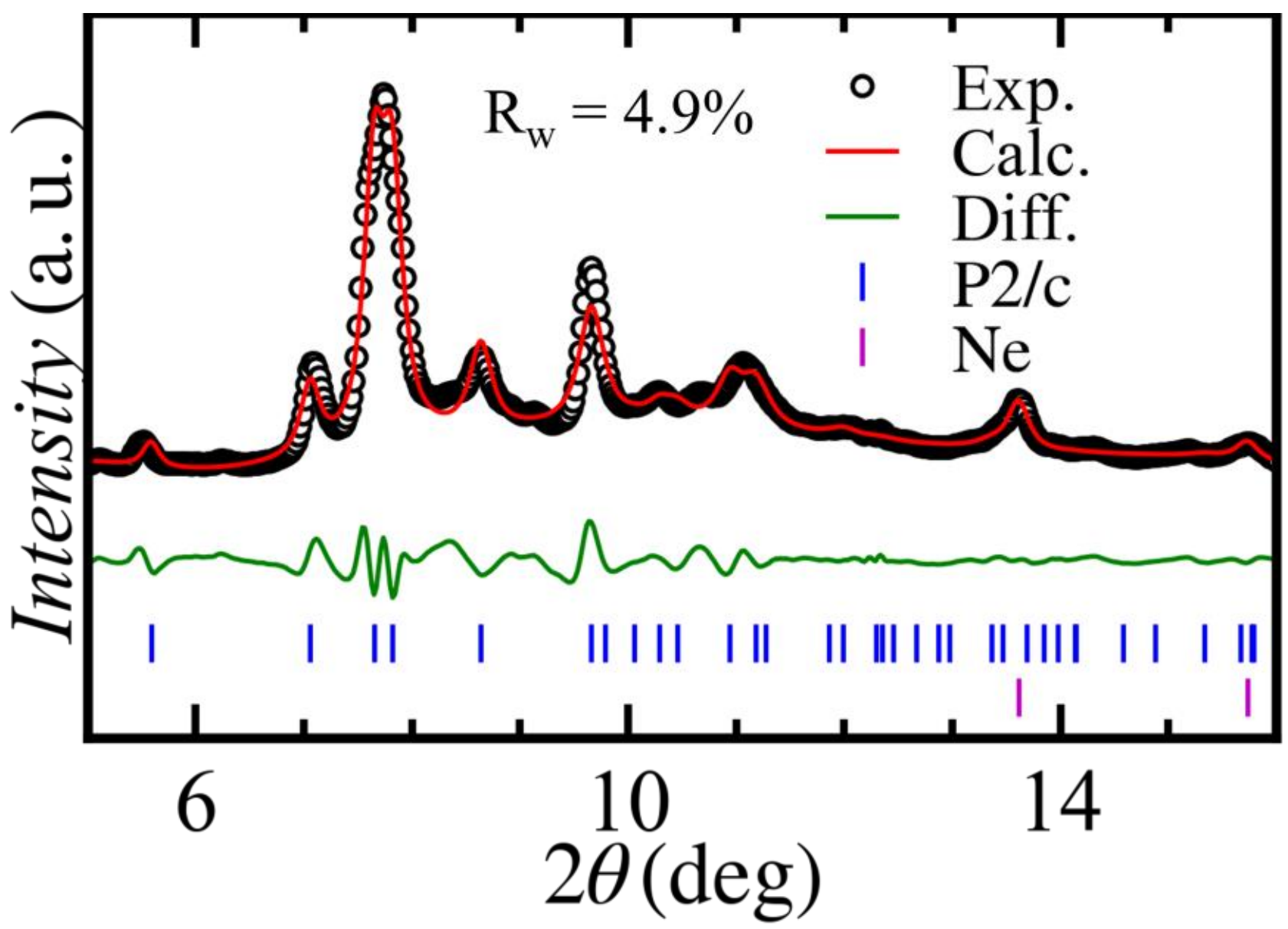

Fig. s4. Le Bail fitting of the 55 GPa diffraction pattern from sample #2 using the *P*2/*c* (#13) structural model for $ReO_3$.

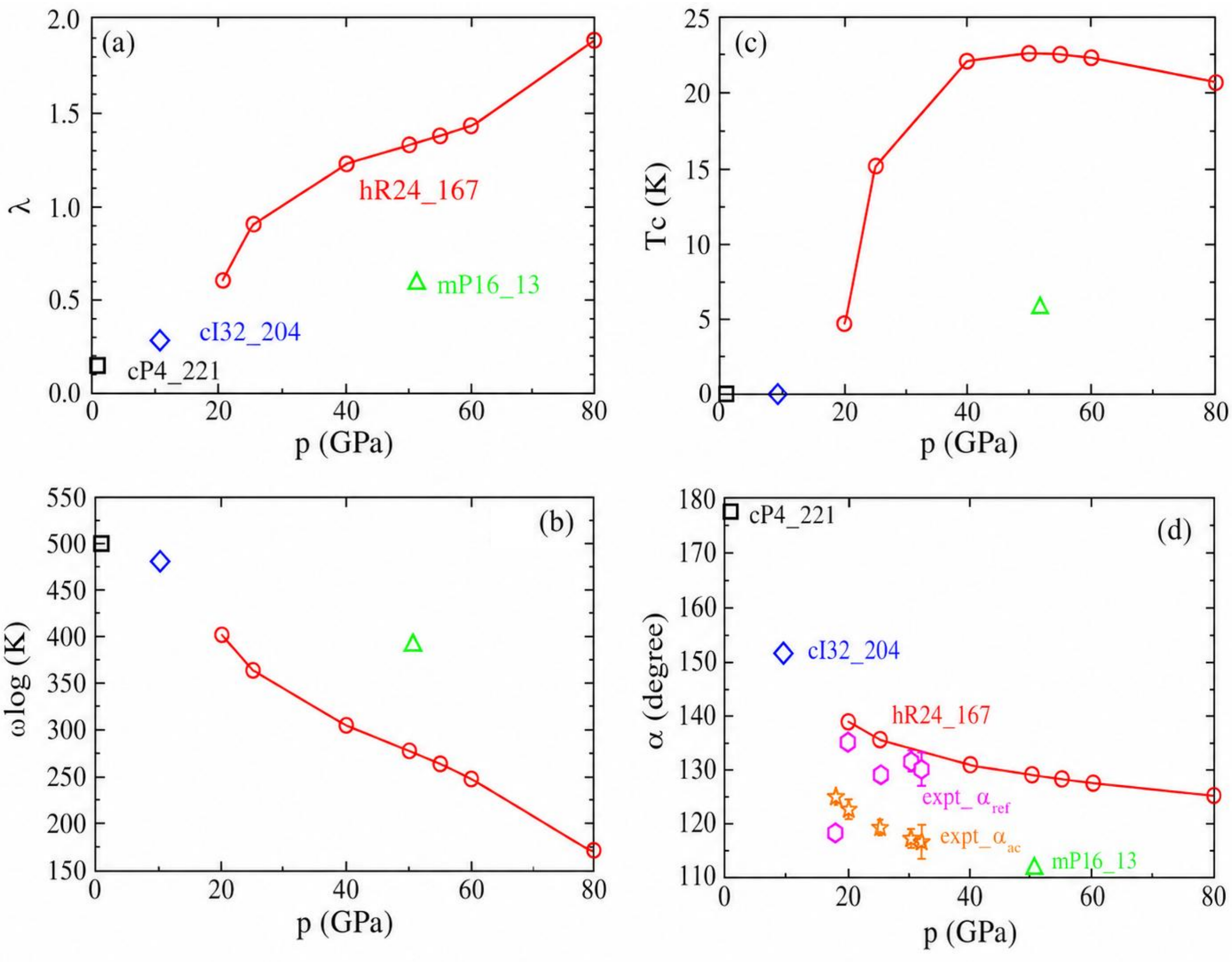

Fig. s5 Pressure dependences of the calculated (a) electron–phonon coupling constant $\lambda$, (b) logarithmic average phonon frequency $\omega_{log}$, (c) superconducting transition temperature $T_c$, and (d) Re-O-Re bond angle α for the different $ReO_3$ phases. The experimental $\alpha_{ref}$ and $\alpha_{ac}$ values are also shown in (d) for comparison. Red lines connect the calculated hR24/ $R\bar{3}c$ data as guides to the eye.

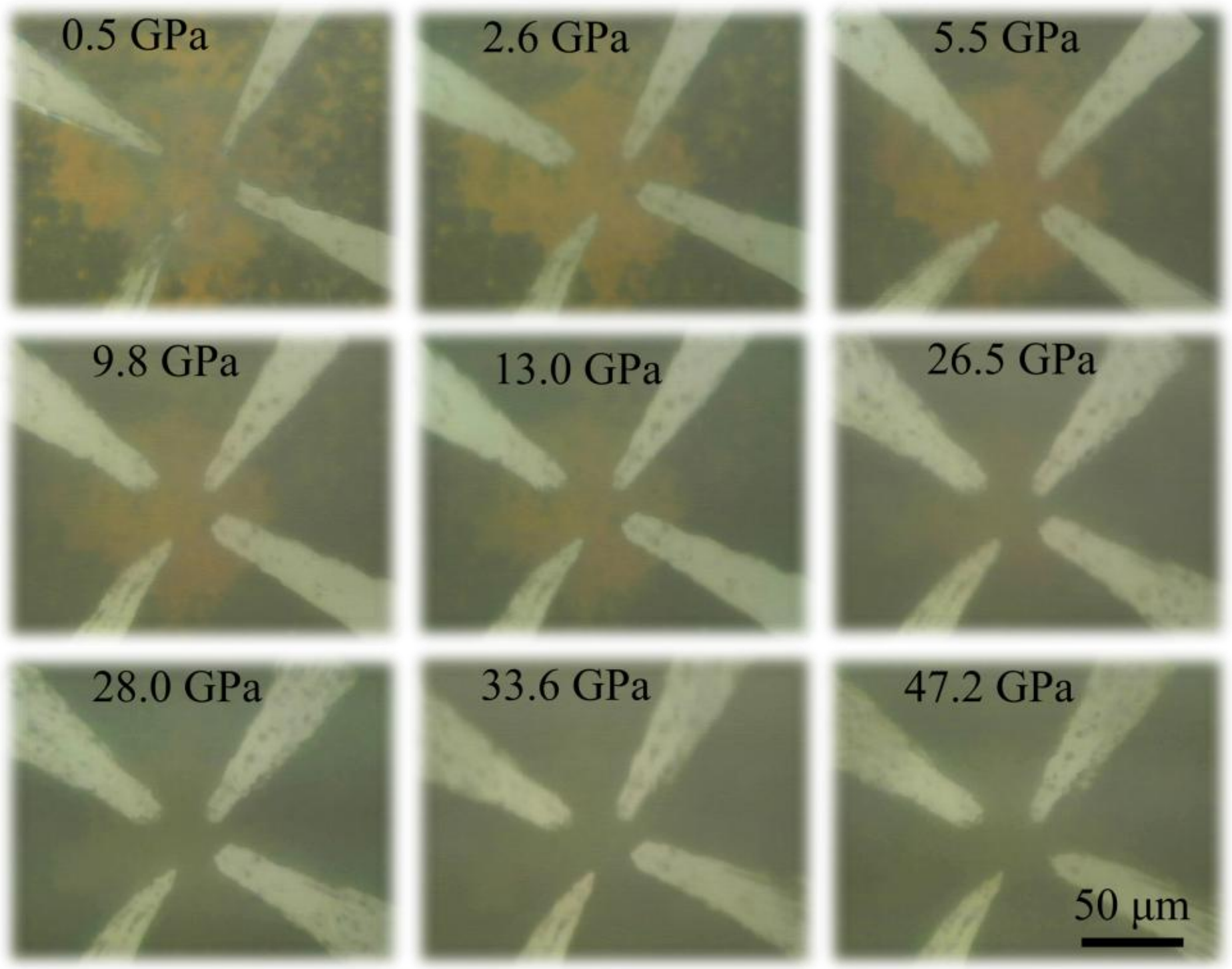

Fig. s6 Selected optical micrographs of the $ReO_3$ powder sample collected at different pressures. The nine photos share the same scale bar shown in the bottom right corner.

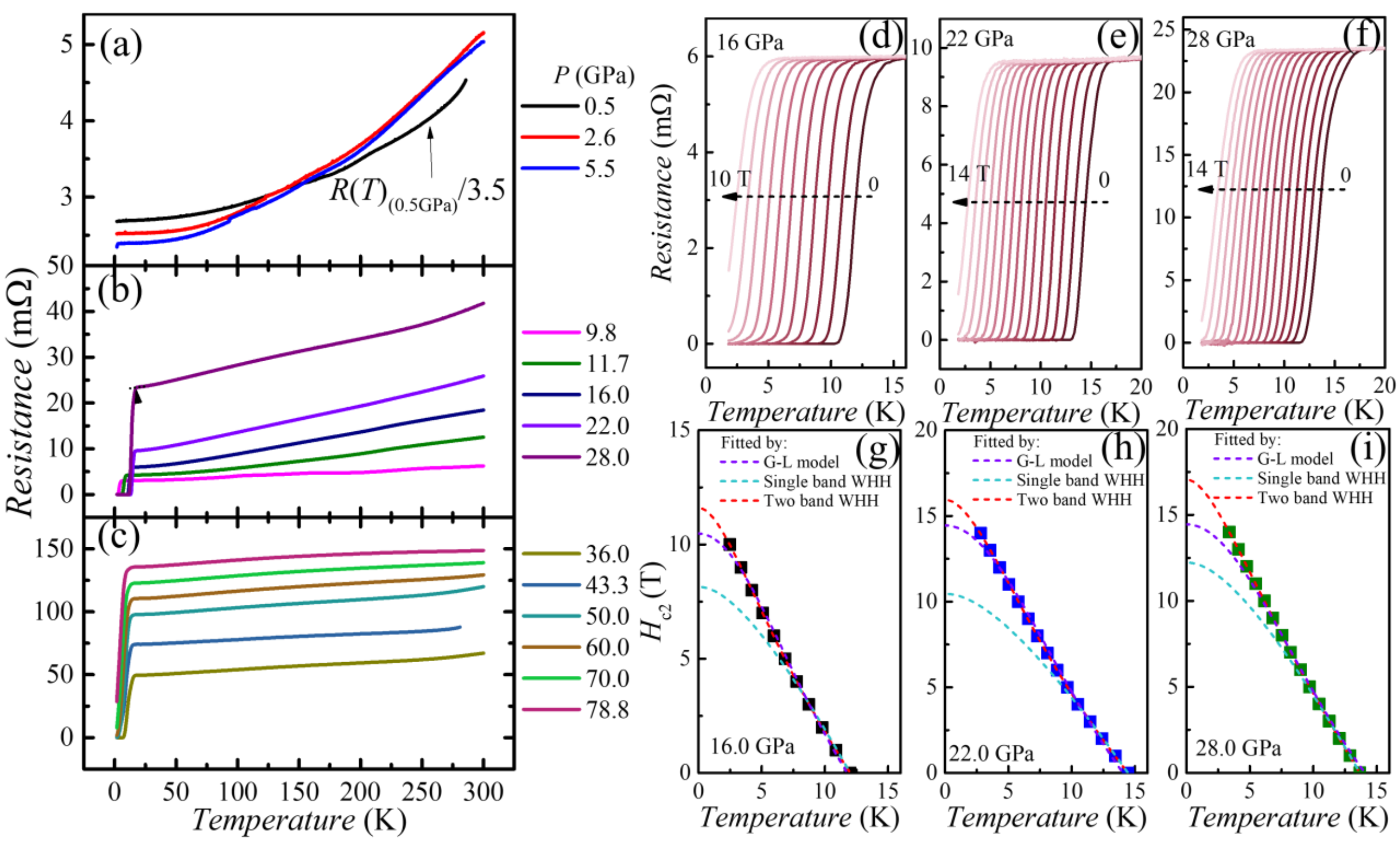

Fig. s7 (a)-(c) at various pressures from 0.5 GPa to 78.8 GPa in full temperature range. (d) - (f) $R(T)$ at different magnetic fields for $P$ = 16.0 GPa, 22.0 GPa and 28 GPa. (g) - (i) The corresponding $R(T)$ plots at different magnetic fields $P$ = 16.0 GPa, 22.0 GPa and 28 GPa.

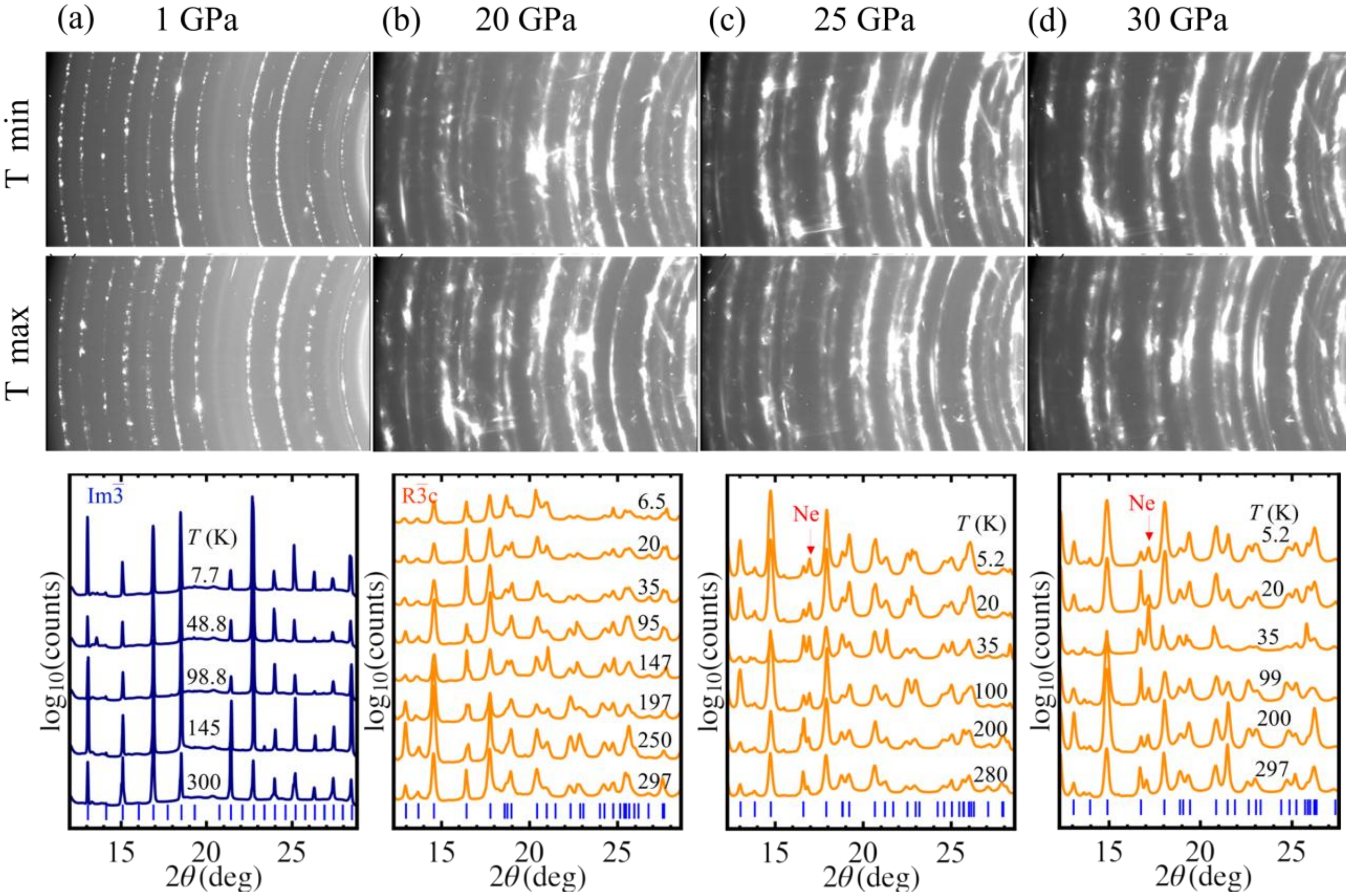

Fig. s8 Two-dimensional (2D) diffraction images and powder diffraction patterns collected at various temperatures under selected pressures: (a) 1 GPa with the sample in the $Im\bar{3}$ phase; (b) 20 GPa, (c) 25 GPa, and (d) 30 GPa with the sample in the $R\bar{3}c$ phase. In each panel, the top row displays the 2D image acquired at the minimum temperature, the middle row shows the 2D image collected at the maximum temperature, and the bottom row presents the temperature-dependent diffraction patterns integrated from the corresponding 2D images. Red arrows indicate additional diffraction peaks arising from solidified neon (pressure transmitting medium). The synchrotron X-ray wavelength is 0.4875 Å. All patterns were refined using Le Bail fitting, yielding $R_w$ in the range of approximately 3–7%.

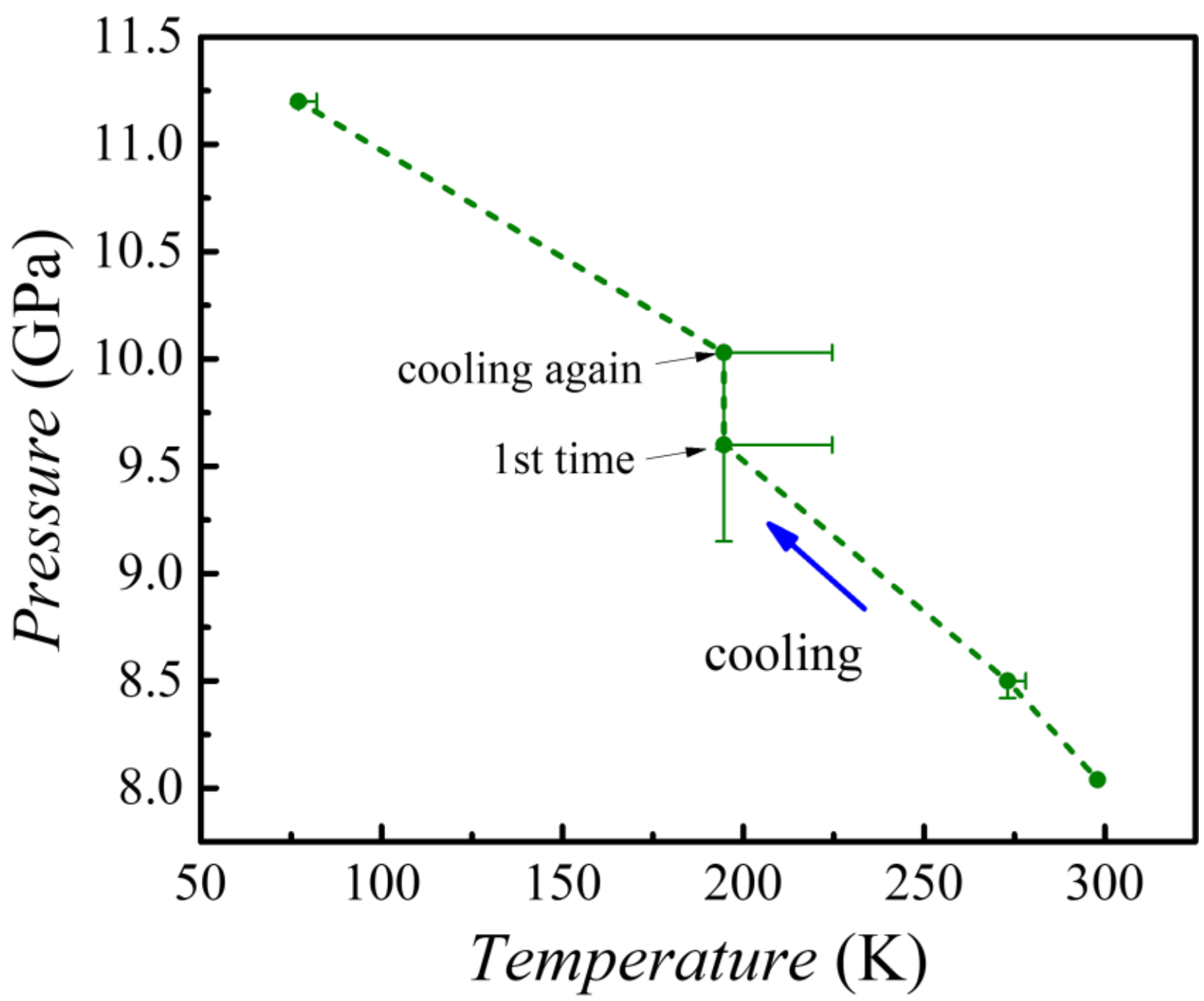

Fig. s9 Pressure-temperature test in the Cu-Be DAC used in the electrical resistance experiment. The temperature error bars are from the temperature changes during pressure measurements and the pressure uncertainties are estimated from the corresponding temperature changes.

Table S1 Crystal structure of $ReO_3$ at 25.2 GPa (room temperature) determined from Rietveld refinement of run #1 data (PTM: M-E mixture). Uncertainties are given in parentheses. $R_w = 4.9\%$.

| Crystal structure | |
|---|---|
| Space group | R-3c (167) |
| a (Å) | 4.613 (23) |
| b (Å) | 4.613 (23) |
| c (Å) | 13.100 (7) |
| Volume (Å$^3$) | 241.39 (8) |
| Z | 6 |

| Atomic coordinates | | | | |
|---|---|---|---|---|
| Atoms | Wyck. | x | y | z |
| Re | 6b | 0 | 0 | 0.5 |
| O | 18e | 0.33333 | 0. 345 (6) | 0.41667 |

Table S2 The possible indexing for the 55 GPa PXRD spectrum for run #2.

| **Refined (Le Bail fitting of phase + Ne peaks)** | | | | | | | |
|---|---|---|---|---|---|---|---|
| **Structure** | **Space group** | **ISODISTORT** | **a (Å)** | **c (Å)** | **V (Å³)** | **$R_w$ (%)** | **red $X^2$** |
| Rhombohedral | $R32$ | T3 | 4.642(8) | 26.05(3) | 486.2(7) | 8.2 | 0.74 |
| Rhombohedral | $R\bar{3}c$ | F1- | 8.81(1) | 13.83(1) | 929(2) | 7.9 | 0.68 |
| **Rhombohedral** | $\boldsymbol{R32}$ | **L1** | **9.876(5)** | **26.632(6)** | **2249(1)** | **3.6** | **0.14** |
| Hexagonal | $P6/mmm$ | - | 11.1286 | 4.15136 | 445.25 | 7.1 | 0.53 |
| Hexagonal | $P63/mmc$ | - | 7.2594 | 12.8789 | 587.8 | 6.3 | 0.43 |
| Tetragonal | $I41/amd$ | - | 10.7041 | 12.7439 | 1460.17 | 6.9 | 0.51 |
| Tetragonal | $P42/ncm$ | - | 7.5772 | 9.0132 | 516.8 | 8.6 | 0.79 |
| Tetragonal | $P42/mnm$ | - | 14.2261 | 6.1442 | 1234.5 | 6.04 | 0.39 |
| **Monoclinic** | $\boldsymbol{P2/c}$ | **(β=116.72(7)°)** | **6.37(4)** | **4.3970(27)** | **5.029(26)** | **4.9** | |